\documentclass[aps,preprint]{revtex4}
\usepackage{amsmath}
\usepackage[english]{babel}
\usepackage{amssymb}
\usepackage{graphicx}
\usepackage{epsfig}
\usepackage{bm}
\usepackage{hyperref}
\usepackage{color}

\def\trmin{{\rm tr}}
\def\mf{{\mbox{\tiny MF}}}
\DeclareMathOperator*{\sumint}{%
\mathchoice%
  {\ooalign{$\displaystyle\sum$\cr\hidewidth$\displaystyle\int$\hidewidth\cr}}
  {\ooalign{\raisebox{.14\height}{\scalebox{.7}{$\textstyle\sum$}}\cr\hidewidth$\textstyle\int$\hidewidth\cr}}
  {\ooalign{\raisebox{.2\height}{\scalebox{.6}{$\scriptstyle\sum$}}\cr$\scriptstyle\int$\cr}}
  {\ooalign{\raisebox{.2\height}{\scalebox{.6}{$\scriptstyle\sum$}}\cr$\scriptstyle\int$\cr}}
}

\begin{document}

\title{\sc\Large{Neutral and charged pion properties under strong magnetic fields in the NJL model} \vspace*{0.5cm}}

\author{M. Coppola$^{a,b}$, D. Gomez Dumm$^{c}$, S. Noguera$^{d}$ and N.N.\ Scoccola$^{a,b}$ \vspace*{0.1cm}}

\affiliation{\small $^a$ CONICET, Rivadavia 1917, 1033 Buenos Aires, Argentina}
\affiliation{$^b$ Physics Department, Comisi\'{o}n Nacional de Energ\'{\i}a At\'{o}mica, Avenue Libertador 8250, 1429 Buenos Aires, Argentina}
\affiliation{$^c$ IFLP, CONICET---Departamento de F\'{\i}sica, Facultad de Ciencias Exactas, Universidad Nacional de La Plata, C.C. 67, 1900 La Plata, Argentina}
\affiliation{$^d$ Departamento de F\'{\i}sica Te\'{o}rica and IFIC, Centro Mixto
Universidad de Valencia-CSIC, E-46100 Burjassot (Valencia), Spain \vspace*{2cm}}

\begin{abstract}
In the framework of the Nambu--Jona-Lasino (NJL) model, we study the effect
of an intense external uniform magnetic field on neutral and charged pion
masses and decay form factors. In particular, the treatment of charged pions
is carried out on the basis of the Ritus eigenfunction approach to
magnetized relativistic systems. Our analysis shows that in the presence of
the magnetic field three and four nonvanishing pion-to-vacuum hadronic form
factors can be obtained for the case of the neutral and charged pions,
respectively. As expected, it is seen that for nonzero magnetic field
the $\pi^0$ meson can still be treated as a pseudo Nambu-Goldstone boson,
and consequently the corresponding form factors are shown to satisfy various
chiral relations. For definite parametrizations of the model, numerical
results for $\pi^0$ and $\pi^\pm$ masses and decay constants are obtained
and compared with previous calculations given in the literature.
\end{abstract}


\maketitle

\renewcommand{\thefootnote}{\arabic{footnote}}
\setcounter{footnote}{0}

\section{Introduction}

In recent years a significant effort has been devoted to the study of the
properties of strongly interacting matter under the influence of strong
magnetic fields (see
e.g.~\cite{Kharzeev:2012ph,Andersen:2014xxa,Miransky:2015ava} and
references therein). This is mostly motivated by the realization that large
magnetic fields might play an important role in the physics of the early
Universe~\cite{Grasso:2000wj}, in the analysis of high energy noncentral
heavy ion collisions~\cite{HIC}, and in the description of physical systems
such as magnetars~\cite{duncan}. From the theoretical point of view,
addressing this subject requires one to deal with quantum chromodynamics (QCD)
in nonperturbative regimes. Therefore, existing analyses are based either in
the predictions of effective models or in the results obtained through
lattice QCD (LQCD) calculations. In this work we focus on the effect of an
intense external magnetic field on $\pi$ meson properties. This issue has
been studied in the last years following various theoretical approaches for
low energy QCD, such as Nambu-Jona-Lasinio (NJL)-like
models~\cite{Fayazbakhsh:2013cha,Fayazbakhsh:2012vr,Avancini:2015ady,Zhang:2016qrl,Avancini:2016fgq,Mao:2017wmq,GomezDumm:2017jij,Wang:2017vtn,Liu:2018zag,Coppola:2018vkw,Mao:2018dqe,Avancini:2018svs},
quark-meson models~\cite{Kamikado:2013pya,Ayala:2018zat}, chiral
perturbation theory
(ChPT)~\cite{Andersen:2012zc,Agasian:2001ym,Colucci:2013zoa}, path integral
Hamiltonians~\cite{Orlovsky:2013wjd,Andreichikov:2016ayj}, effective chiral
confinement Lagrangian approach (ECCL)~\cite{Simonov:2015xta,Andreichikov:2018wrc},
QCD sum rules (SRQCD)~\cite{Dominguez:2018njv}, etc. In addition, results for the
light meson spectrum in the presence of background magnetic fields have been
recently obtained from LQCD
calculations~\cite{Bali:2011qj,Hidaka:2012mz,Luschevskaya:2014lga,Bali:2015vua,Bali:2017ian}.

In the framework of the NJL model, mesons are usually described as quantum
fluctuations in the random phase approximation
(RPA)~\cite{Vogl:1991qt,Klevansky:1992qe,Hatsuda:1994pi}, i.e., they are
introduced via the summation of an infinite number of quark loops. In the
presence of a magnetic field $\vec B$, the calculation of these loops
requires some care due to the appearance of Schwinger
phases~\cite{Schwinger:1951nm} associated with quark propagators. For the
neutral pion these phases cancel out, and as a consequence the usual
momentum basis can be used to diagonalize the corresponding polarization
function~\cite{Fayazbakhsh:2013cha,Fayazbakhsh:2012vr,Avancini:2015ady,Avancini:2016fgq,Mao:2017wmq}.
On the other hand, for charged pions the Schwinger phases do not cancel,
leading to a breakdown of translational invariance that prevents one from proceeding
as in the $\pi^0$ case. In this situation, some existing
calculations~\cite{Zhang:2016qrl,Liu:2018zag} just neglect Schwinger phases,
considering only the translational invariant part of the quark propagators.
Recently~\cite{Coppola:2018vkw}, a method was proposed in order
to fully take into account the translational-breaking effects introduced by
the Schwinger phases in the calculation of charged meson masses within the
RPA. This method, based on the Ritus eigenfunction
approach~\cite{Ritus:1978cj} to magnetized relativistic systems, allows one to
diagonalize the charged pion polarization function for the obtention of
the corresponding meson masses. In addition, the analysis in
Ref.~\cite{Coppola:2018vkw} considers a regularization procedure in which
only the vacuum contributions to different quantities at zero external
magnetic field are regularized. This scheme, that goes under the name of
``magnetic field independent regularization'' (MFIR), has been shown to
provide more reliable predictions in comparison with other regularization
methods often used in the literature~\cite{Avancini:2019wed}.

The scope of the present work is to consider the approach introduced in
Ref.~\cite{Coppola:2018vkw} for the study of pion masses, extending the
calculations to other properties of neutral and charged pions. In
particular, we concentrate in the analysis of the form factors associated
with the pion-to-vacuum matrix elements of the vector and axial vector
hadronic currents. For the case of the $\pi^0$, some
works~\cite{Fayazbakhsh:2013cha,Agasian:2001ym,Andersen:2012zc,
Fayazbakhsh:2012vr,Avancini:2015ady,Zhang:2016qrl,Avancini:2016fgq,
Mao:2017wmq,Kamikado:2013pya,GomezDumm:2017jij,Simonov:2015xta,Andreichikov:2018wrc}
have already considered the $B$ dependence of the decay constant
$f_{\pi^0}^{(A1)}$, which corresponds to the time component of the axial
vector current matrix element. For charged pions, the effect of the magnetic
field has been analyzed in the context of ChPT~\cite{Andersen:2012zc},
ECCL approach~\cite{Simonov:2015xta}, SRQCD~\cite{Dominguez:2018njv} and,
quite recently, through LQCD calculations~\cite{Bali:2018sey}. An
interesting observation made in Ref.~\cite{Fayazbakhsh:2013cha} states that,
due to the explicit breaking of rotational invariance caused by the magnetic
field, one can define two different decay constants. One of them is
associated with the direction parallel to $\vec B$ (and to the time
direction) and the other with the spatial directions perpendicular to $\vec
B$. A further relevant statement has been pointed out in
Ref.~\cite{Bali:2018sey}. In that work it is noted that the presence of the
background magnetic field opens the possibility of a nonzero charged
pion-to-vacuum transition via the vector piece of the electroweak current.
This implies the existence of a further decay constant associated with the
pion-to-vacuum matrix element of the vector current. Furthermore, in a
recent work~\cite{Coppola:2018ygv} we have shown that (i) the
pion-to-vacuum matrix element of the vector current can be nonvanishing even
in the case of the neutral pion, and (ii) for the charged pions there are
in general not two but {\it three} nonvanishing axial decay constants. The
aim of the present paper is to study the behavior of all these form factors
as functions of the magnetic field in the context of the NJL model within
the MFIR regularization scheme.

This work is organized as follows. In Sec.~II we introduce the theoretical
formalism used to obtain the different quantities we are interested in.
Chiral limit relations are addressed in Sec.~III. Then, in Sec.~IV we present and
discuss our numerical results, while in Sec.~V we provide a summary of our
work, together with our main conclusions. We also include
Appendixes A and B to quote some technical details of our calculations.

\section{Theoretical Formalism}

\subsection{Mean field properties and pion masses}

We start by considering the Euclidean Lagrangian density for the NJL
two-flavor model in the presence of an electromagnetic field. One has
\begin{equation}
{\cal L} \ = \ \bar \psi \left(- i\, \rlap/\!D + m_0 \right) \psi - G \left[
(\bar\psi \, \psi)^2 + (\bar\psi\, i\gamma_{5}\vec{\tau}\,\psi) \right]\ ,
\qquad \psi=\left(\begin{array}{c}
\psi_{u}\\
\psi_{d}
\end{array}\right) \ ,
\label{lagrangian}
\end{equation}
where $\tau_i$ are the Pauli matrices and $m_0$ is the
current quark mass, which is assumed to be equal for $u$ and $d$ quarks. The
interaction between the fermions and the electromagnetic field ${\cal
A}_\mu$ is driven by the covariant derivative
\begin{equation}
D_\mu \, = \, \partial_{\mu}-i\,\hat Q \mathcal{A}_{\mu}\ ,
\label{covdev}
\end{equation}
where $\hat Q=\mbox{diag}(q_u,q_d)$, with $q_u=2e/3$ and $q_d = -e/3$, $e$
being the proton electric charge. We will consider the particular case of an
homogenous stationary magnetic field $\vec B$ along the positive 3-axis. Let
us choose the Landau gauge, in which $\mathcal{A}_4 = 0$, $\vec{\mathcal{A}} =
(0,B x_1,0)$.

Since we are interested in studying meson properties, it is convenient to
bosonize the fermionic theory, introducing scalar $\sigma(x)$ and pseudoscalar $\vec{\pi}(x)$
fields and integrating out the fermion fields. The
bosonized Euclidean action can be written as~\cite{Klevansky:1992qe}
\begin{equation}
S_{\mathrm{bos}} \, = \, -\log\det\mathcal{D}+\frac{1}{4G}
\int d^{4}x\
\Big[\sigma(x)\sigma(x)+ \vec{\pi}(x)\cdot\vec{\pi}(x)\Big]\ ,
\label{sbos}
\end{equation}
with
\begin{equation}
\mathcal{D}_{x,x'} \, = \, \delta^{(4)}(x-x')\, \left[-i\,\rlap/\!D + m_0 +
\sigma(x) + i\,\gamma_5\,\vec{\tau}\cdot\vec{\pi}(x)  \right]\ ,
\label{dxx}
\end{equation}
where a direct product to an identity matrix in color space is understood.

We proceed by expanding the bosonized action in powers of the fluctuations
$\delta\sigma(x)$ and $\delta\pi_i(x)$ around the corresponding mean field
(MF) values. As usual, we assume that the field $\sigma(x)$ has a nontrivial
translational invariant mean field value $\bar{\sigma}$, while the vacuum
expectation values of pseudoscalar fields are zero. Thus we write
\begin{equation}
\mathcal{D}_{x,x'} \ = \ \mathcal{D}^{\mbox{\tiny MF}}_{x,x'} + \delta\mathcal{D}_{x,x'}\ .
\label{dxxp}
\end{equation}
The MF piece is flavor diagonal. It can be written as
\begin{equation}
\mathcal{D}^\mf_{x,x'} \ = \ {\rm diag}\big(\mathcal{D}^{\mf,u}_{x,x'}\, ,\,
\mathcal{D}^{\mf,d}_{x,x'}\big)\ ,
\end{equation}
where
\begin{equation}
\mathcal{D}^{\mf,f}_{x,x'} \ = \ \delta^{(4)}(x-x') \left( - i \rlap/\partial
- q_f \, B \, x_1 \, \gamma_2 + m_0 + \bar\sigma \right).
\end{equation}
On the other hand, the second term on the right-hand-side of Eq.~(\ref{dxxp}) is given
by
\begin{equation}
\delta D_{x,x'} \ =\ \delta^{(4)}(x-x') \,
 \begin{pmatrix}
   \delta\sigma(x) + i\gamma_5 \delta\pi_0(x) & \sqrt{2}i\gamma_5\,\delta\pi^+(x) \\
    \sqrt{2}i\gamma_5\,\delta\pi^-(x) & \delta\sigma(x) - i\gamma_5 \delta\pi_0(x)
  \end{pmatrix}\ ,
  \label{DeltaD}
\end{equation}
where $\pi^\pm=\left(\pi_1 \mp i \pi_2\right)/\sqrt{2}$. Replacing in the
bosonized effective action and expanding in powers of the meson fluctuations
around the MF values, we get
\begin{eqnarray}
S_{\mathrm{bos}} \ = \ S^{\mbox{\tiny MF}}_{\mathrm{bos}} \, + \,
S^{\,\mbox{\tiny quad}}_{\mathrm{bos}}\, + \,\dots
\end{eqnarray}
Here, the mean field action per unit volume reads
\begin{equation}
\frac{S^{\mbox{\tiny MF}}_{\mathrm{bos}}}{V^{(4)}} \ = \ \frac{ \bar
\sigma^2}{4 G} - \frac{N_c}{V^{(4)}} \sum_{f=u,d} \int d^4x \, d^4x' \
\trmin_D\, \ln \left(\mathcal{S}^{\mbox{\tiny MF},f}_{x,x'}\right)^{-1} \ ,
\label{seff}
\end{equation}
where $\trmin_D$ stands for the trace in Dirac space. The quadratic
contribution is given by
\begin{equation}
S^{\,\mbox{\tiny quad}}_{\mathrm{bos}}\ = \ \dfrac{1}{2}
\sum_{M=\sigma,\pi^0,\pi^\pm} \int d^4x \, d^4x'\ \delta M
(x)^\ast \left[  \frac{1}{2 G}\; \delta^{(4)}(x-x') - J_M(x,x')
\right] \delta M(x')\ , \label{actionquad}
\end{equation}
where
\begin{eqnarray}
J_{\pi^0} (x,x') &=& N_c \sum_f \trmin_D \bigg[
\mathcal{S}^{\mf,f}_{x,x'} \ \gamma_5 \ \mathcal{S}^{\mf,f}_{x',x}
\ \gamma_5 \ \bigg]\ ,
\nonumber \\
J_{\pi^-} (x,x') &=& 2 N_c  \, \trmin_D \bigg[
\mathcal{S}^{\mf,d}_{x,x'} \ \gamma_5 \ \mathcal{S}^{\mf,u}_{x',x}
\ \gamma_5 \ \bigg]\ ,
\nonumber \\
J_{\pi^+} (x,x') &=& 2 N_c  \, \trmin_D \bigg[
\mathcal{S}^{\mf,u}_{x,x'} \ \gamma_5 \ \mathcal{S}^{\mf,d}_{x',x}
\ \gamma_5 \ \bigg]\ , \label{jotas}
\end{eqnarray}
while the expression for $J_\sigma$ is obtained from that of $J_{\pi^0}$
just replacing both $\gamma_5$ matrices for unit matrices. In these
expressions we have introduced the mean field quark propagators
$\mathcal{S}^{\mf,f}_{x,x'} = \big( \mathcal{D}^{\mf,f}_{x,x'} \big)^{-1}$.
As is well known, their explicit form can be written in different
ways~\cite{Andersen:2014xxa,Miransky:2015ava}. For convenience we take the
form in which $\mathcal{S}^{\mf,f}_{x,x'}$ is given by a product of a phase
factor and a translational invariant function, namely
\begin{equation}
S^{\mf,f}_{x,x'} \ = \ e^{i\Phi_f(x,x')}\,\int_p e^{i\, p\, (x-x')}\, \tilde S_p^f\
,
\label{sfx}
\end{equation}
where $\Phi_f(x,x')= q_f B (x_1+x_1')(x_2-x_2')/2$ is the
so-called Schwinger phase. We have introduced here the shorthand notation
\begin{equation}
\int_{p}\ \equiv \ \int \dfrac{d^4 p}{(2\pi)^4}\ .
\label{notation1}
\end{equation}
We express $\tilde S_p^f$ in the Schwinger form~\cite{Andersen:2014xxa,Miransky:2015ava}
\begin{equation}
\tilde S_p^f \ = \ \int_0^\infty \!d\tau\,
\exp\big[-\tau \phi_f(\tau,p)\big]\,
\Big[\big(M-p_\parallel \gamma_\parallel \big)
\, \big(1+i s_f \,\gamma_1 \gamma_2\, \tanh(\tau B_f)\big) -
\dfrac{p_\perp \gamma_\perp}{\cosh^2(\tau B_f)} \Big] \ ,
\label{sfp_schw}
\end{equation}
where we have used the following definitions. The ``perpendicular'' and
``parallel'' gamma matrices are collected in vectors $\gamma_\perp =
(\gamma_1,\gamma_2)$ and $\gamma_\parallel = (\gamma_3,\gamma_4)$.
Similarly, $p_\perp = (p_1,p_2)$ and $p_\parallel = (p_3,p_4)$. Note that in
our convention $\{\gamma_\mu,\gamma_\nu\}=-2 \delta_{\mu\nu}$. The quark
effective mass $M$ is given by $M=m_0+\bar\sigma$, and we have used the
notation $s_f = {\rm sign} (q_f B)$ and $B_f=|q_fB|$. Finally, we have
defined
\begin{equation}
\phi_f(\tau,p)\ = \ M^2 + p_\parallel^2 +
\dfrac{\tanh(\tau B_f)}{\tau B_f}\; p_\perp^2\ .
\end{equation}
Notice that the integral in Eq.~(\ref{sfp_schw}) is divergent and has to be
properly regularized, as we discuss below.

Replacing the above expression for the quark propagator in Eq.~(\ref{seff})
and minimizing with respect to $M$ we obtain the gap equation
\begin{equation}
M \ = \ m_0 + 4 G M N_c\, I\ ,
\label{gapeq}
\end{equation}
where $I$ is a divergent integral.
To regularize it we use here the magnetic field independent
regularization scheme~\cite{Menezes:2008qt,Allen:2015paa}. That is, we
subtract from $I$ the unregulated integral in the $B=0$ limit, $I_{B=0}$,
and then we add it in a regulated form $I^{\rm (reg)}_{B=0}$. Thus, we have
\begin{equation}
I^{\rm (reg)} \ = \ I^{\rm (reg)}_{B=0} + I^{\rm (mag)}\
,
\label{ireg}
\end{equation}
where $I^{\rm (mag)}$ is a finite, magnetic field dependent contribution given by
\begin{eqnarray}
I^{\rm (mag)} & = &  \frac{1}{8 \pi^2} \sum_f  \int_0^\infty\!
d\tau\; \frac{\exp(-\tau M^2)}{\tau^2}\; \Big[  \tau B_f \coth( \tau B_f ) -1
\Big]\nonumber \\
& = & \frac{M^2}{8\pi^2} \sum_f \left[ \frac{ \ln \Gamma(x_f)}{x_f} -
\frac{\ln 2\pi}{2x_f} + 1 - \left( 1 - \frac{1}{2x_f}\right) \ln x_f
\right]\ ,
\label{imag}
\end{eqnarray}
with $x_f = M^2/(2 B_f)$.  On the other hand, the
regulated piece $I^{\rm (reg)}_{B=0}$ does depend on the regularization
prescription. Choosing the standard procedure in which one introduces a 3D
momentum cutoff $\Lambda$, we get the well-known result
\cite{Klevansky:1992qe}
\begin{equation}
I^{\rm (reg)}_{B=0} \ = \ I_1 \equiv \ \frac{1}{2 \pi^2} \left[
\Lambda\,\sqrt{ \Lambda^2 + M^2 } + M^2\;
\ln{\left( \frac{ M}{\Lambda + \sqrt{\Lambda^2 + M^2}}\right)}\right]\ .
\label{I13d}
\end{equation}

\hfill

For the reader's convenience, in what remains of this subsection we review
the procedure followed in Ref.~\cite{Coppola:2018vkw} to determine the pion
masses. We start by the simpler case of the neutral pion $\pi^0$. In this
case the contributions of Schwinger phases associated to the quark
propagators cancel out. Therefore, the polarization function depends only on
the difference $x-x'$ (i.e., it is translational invariant), which leads to
the conservation of $\pi^0$ momentum. If we take now the Fourier transform
of $\pi^0$ fields to the momentum basis, the corresponding transform of the
polarization function will be diagonal in $q,q'$ momentum space. Thus, the
$\pi^0$ contribution to the quadratic action in the momentum basis can be
written as
\begin{equation}
S^{\,\mbox{\tiny quad}}_{\pi^0} \ = \ \dfrac{1}{2} \int_{q} \
\delta\pi^0(-q) \, \left[\frac{1}{2 G} -
J_{\pi^0}(q_\perp^2,q_\parallel^2) \right] \delta\pi^0(q)\ ,
\label{actionquadpi0p}
\end{equation}
where
\begin{equation}
J_{\pi^0}(q_\perp^2,q_\parallel^2) \ = \ N_c \sum_f \int_{p,v} (2\pi)^4 \delta^{(4)}(v-q)
\ \trmin_D\, (\tilde S_{p^+}^f \, \gamma_5 \, \tilde S_{p^-}^f \,
\gamma_5)\ , \label{J0q}
\end{equation}
with $p^\pm = p \pm v/2$. Replacing Eq.~(\ref{sfp_schw}) into
Eq.~(\ref{J0q}) and using the results in Appendixes A and B one finds
\begin{eqnarray}
J_{\pi^0}(q_\perp^2,q_\parallel^2) &=& \dfrac{N_c}{4\pi^2} \sum_f\, B_f \int_0^\infty dz \int_{0}^1 dy
\, \exp \left\{-z\left[M^2+y(1-y)\,q_\parallel^2\right]\right\} \, \times \nonumber \\
&& \exp\left[-\dfrac{q_\perp^2}{B_f} \,\gamma_f(y,z) \right]
\bigg\{ \Big[M^2+\dfrac{1}{z}-y(1-y)\,q_\parallel^2 \Big] \coth(zB_f) \, + \nonumber \\
&& \dfrac{B_f}{\sinh^2(zB_f)} \left[1 - \dfrac{q_\perp^2}{B_f}\,\gamma_f(y,z) \right] \bigg\}\ ,
\label{J0B}
\end{eqnarray}
where
\begin{equation}
\gamma_f(y,z) \ = \ \dfrac{\sinh(zyB_f)\sinh[(1-y)zB_f]}{\sinh(zB_f)} \ .
\label{gamma_f}
\end{equation}
As usual, here we have used the changes of variables $\tau = y z$ and
$\tau'=(1-y)z$, $\tau$ and $\tau'$ being the integration parameters
associated with the quark propagators as in Eq.~(\ref{sfp_schw}).

As done at the MF level, we regularize the integral in Eq.~(\ref{J0B}) using
the MFIR scheme. That is, we subtract the corresponding unregulated
contribution in the $B=0$ limit, given by
\begin{equation}
J_{\pi,B=0} (q^2) \ = \ \dfrac{N_c}{2\pi^2} \int_0^\infty
\dfrac{dz}{z} \int_{0}^1 dy \,
\exp \left\{-z\left[M^2+y(1-y)\,q^2 \right]\right\} \left[M^2+\dfrac{2}{z}-y(1-y)\,q^2 \right]
\ , \label{TPFB0}
\end{equation}
and add it in a regularized form $J_{\pi,B=0}^{\rm (reg)} (q^2)$.
The regularized polarization function is then given by
\begin{equation}
J^{\rm (reg)}_{\pi^0}(q_\perp^2,q_\parallel^2) \ = \ J^{\rm
(reg)}_{\pi,B=0} (q^2) + J^{\rm
(mag)}_{\pi^0}(q_\perp^2,q_\parallel^2)\ ,
\label{jregmag}
\end{equation}
where $J^{\rm (mag)}_{\pi^0}(q_\perp^2,q_\parallel^2) =
J_{\pi^0}(q_\perp^2,q_\parallel^2) - J_{\pi,B=0} (q_\perp^2+q_\parallel^2)$.
To get $J^{\rm (reg)}_{\pi,B=0} (q^2)$ we use the 3D momentum cutoff
scheme, as in the case of the gap equation. One has in this way
\begin{equation}
J^{\rm (reg)}_{\pi,B=0}(q^2) \ = \ 2 N_c \left[ I_1 + q^2 I_2(q^2) \right]\ ,
\label{J+0reg}
\end{equation}
where $I_1$ is given by Eq.~(\ref{I13d}), while
\begin{eqnarray}
\hspace{-0.5cm} I_2(q^2) &=& \frac{1}{4 \pi^2} \int_0^1 dy \left[
\frac{\Lambda}{\sqrt{\Lambda^2+ M^2 + y(1-y) q^2}}\, +\, \ln{ \frac{
{\sqrt{M^2 + y(1-y) q^2}}}{\Lambda + \sqrt{\Lambda^2 +
M^2 + y(1-y) q^2}}} \right] .
\label{I23d}
\end{eqnarray}
Choosing the frame in which the $\pi^0$ meson is at rest, its mass
can be obtained by solving the equation
\begin{equation}
\frac{1}{2 G} - J^{\rm (reg)}_{\pi^0}(0,-m_{\pi^0}^2) \ = \ 0\ .
\end{equation}

\hfill

Let us now discuss the case of charged pions. For definiteness we
consider the $\pi^-$ meson, although a similar analysis, leading to the same
expression for the $B$-dependent mass, can be carried out for the $\pi^+$.
As in the case of the $\pi^0$, we start by replacing Eq.~(\ref{sfx}) in the
expression of the corresponding polarization function in Eq.~(\ref{jotas}).
We get
\begin{eqnarray}
J_{\pi^-}(x,x') \ = \ 2N_c \int_{p\,v} \trmin_D
(\tilde S_{p^+}^d \, \gamma_5 \, \tilde S_{p^-}^u \, \gamma_5)\,
e^{i\Phi_d(x,x')}\, e^{i\Phi_u(x',x)}\, e^{i v (x-x')}\  .
\label{J+q1}
\end{eqnarray}
where once again we define $p^\pm = p \pm v/2$. Contrary to the $\pi^0$
case, here the Schwinger phases do not cancel, due to their different quark
flavors. Therefore, this polarization function is not translational
invariant, and consequently it will not become diagonal when transformed to
the momentum basis. Therefore, we expand the charged pion field as
\begin{equation}
\pi^-(x) \ = \ \sumint_{\bar q}\ \mathbb{F}_{\bar q}^-(x) \,
\pi_{\bar q}^- \ ,
\label{Ritus}
\end{equation}
where we have used the shorthand notation
\begin{equation}
\bar q \equiv (\breve{q},q_4)\ , \qquad
\breve q \equiv (k,q_2,q_3)
\ ,\qquad \sumint_{\bar q}\ \equiv \ \dfrac{1}{2\pi}\sum_{k=0}^\infty \int_{q_2\,q_3\,q_4}\ .
\label{notation2}
\end{equation}
The functions $\mathbb{F}_{\bar q}^\pm(x)$ are given by
\begin{equation}
\mathbb{F}_{\bar q}^\pm(x) \ = \ N_k \, e^{i ( q_2 x_2 + q_3 x_3 + q_4 x_4)}
\, D_k(\rho_{\pm})\ ,
\label{Fq}
\end{equation}
where $D_k(x)$ are the cylindrical parabolic functions. We have used the
definitions $N_k= (4\pi B_e)^{1/4}/\sqrt{k!}$ and $\rho_\pm =
\sqrt{2B_e}\,x_1-s_\pm\sqrt{2/B_e}\,q_2$, where $B_e = |q_{\pi^\pm} B| =
|eB|$ and $s_\pm= \mathrm{sign}(q_{\pi^\pm} B)$, with $q_{\pi^\pm} = \pm(q_u - q_d)
=\pm e$. For the $\pi^-$ one has
\begin{equation}
S^{\,\mbox{\tiny quad}}_{\pi^-} \ = \ \dfrac{1}{2} \sumint_{\bar q', \bar q}
\ (\delta\pi_{\bar q}^-)^\ast \Big( \frac{1}{2G}\, \hat \delta_{\bar q \bar q'} -
J^-_{\bar q\bar q'}\Big) \delta\pi_{\bar q'}^- \ ,
\label{actionquadTPF}
\end{equation}
where
\begin{equation}
\hat \delta_{\bar q \bar q'} = (2\pi)^4
\delta_{kk'} \delta(q_2-q_2') \delta(q_3-q_3') \delta(q_4-q_4')\ .
\end{equation}
and
\begin{equation}
J^-_{\bar q\bar q'} \ = \ 2N_c\int_{p\,v} \!\!\trmin_D
\left[\tilde{S}_{p^+}^d\gamma_5\, \tilde{S}_{p^-}^u\gamma_5\, \right]\int d^4x \, d^4x'\,  e^{i\Phi_d(x,x')}
e^{i\Phi_u(x',x)} e^{i v(x-x')} \, \mathbb{F}_{\bar q}^-(x)^\ast\, \mathbb{F}_{\bar
q'}^-(x')\ .
\label{Fmu+1}
\end{equation}

Integrating over $x'$ in Eq.~(\ref{Fmu+1}) one obtains
\begin{eqnarray}
J^-_{\bar q\bar q'} & = & \dfrac{8\pi N_c}{B_e}
\int d^4x \; \mathbb{F}_{\bar q}^-(x)^\ast\; e^{i(q'_2x_2+q'_\parallel x_\parallel)}
\int_{p\,v} \!\!(2\pi)^2\,\delta^{(2)}(v_\parallel-q_\parallel')\,
\trmin_D\left[\tilde{S}_{p^+}^d \gamma_5\, \tilde{S}_{p^-}^u \gamma_5\,
\right] \times \nonumber \\
& & N_{k'}\,e^{i v_1(x_1-x'_1)}\,D_{k'}
(\sqrt{2B_e}\,x_1'-s_-\sqrt{2/B_e}\,q'_2)\Big|_{x_1' =
-x_1+2s_-(q'_2-v_2)/B_e}\ .
\label{jota+}
\end{eqnarray}
The integrals over the loop momenta $p$ and $v$ can be evaluated using the
results in Appendixes A and B. It can be shown that the polarization function is
diagonal in the chosen basis. One has
\begin{equation}
J^-_{\bar q\bar q'} \ = \ \int d^4x \;
\mathbb{F}_{\bar q}^-(x)^\ast J_{\pi^-}(k,\Pi^2)\; \mathbb{F}_{\bar q'}^-(x)
\ = \ \hat \delta_{\bar q \bar q'} \; J_{\pi^-}(k,\Pi^2)\ ,
\end{equation}
where $\Pi^2 = (2k+1)B_e+q_\parallel^2$, and
\begin{eqnarray}
J_{\pi^-}(k,\Pi^2) &=& \dfrac{N_c}{2\pi^2} \int_0^\infty\! dz
\int_{0}^1 dy \ \exp\big[-zM^2-zy(1-y)(\Pi^2-(2k+1)\, B_e)\big]\ \times
\nonumber \\
&& \dfrac{\alpha_-^k}{\alpha_+^{k+1}}\; \bigg\{\Big[M^2+\dfrac{1}{z}-y(1-y)(\Pi^2-(2k+1)\, B_e)
\Big](1-t_u \,t_d) \,+
\nonumber \\
&& \dfrac{(1-t_u^2)(1-t_d^2)}{\alpha_+\,\alpha_-}\, \Big[ \alpha_- + (\alpha_- - \alpha_+)\,k \Big] \bigg\}\ .
\label{J-B}
\end{eqnarray}
Here we have introduced the definitions $t_u=\tanh (B_u y z)$, $t_d=\tanh
[B_d (1-y) z]$ and $\alpha_\pm = (B_d t_u+B_u t_d \pm B_e \,t_ut_d)/(B_u
B_d)$. For the $\pi^+$, one can show that $J_{\pi^+}(k,\Pi^2) =
J_{\pi^-}(k,\Pi^2)$.

As in the case of the neutral pion, the polarization function in
Eq.~(\ref{J-B}) turns out to be divergent and has to be regularized. Once
again, this can be done within the MFIR scheme. However, due to quantization
in the 1-2 plane this requires some care, viz.~the subtraction of the $B=0$
contribution to the polarization function has to be carried out once the
latter has been written in terms of the squared canonical momentum $\Pi^2$,
as in Eq.~(\ref{J-B}). Thus, the regularized $\pi^-$ polarization function
is given by
\begin{equation}
J_{\pi^-}^{{\rm (reg)}}(k,\Pi^2) \ = \ \, J^{\rm
(reg)}_{\pi,B=0}(\Pi^2) \, +  \, J_{\pi^-}^{{\rm (mag)}}(k,\Pi^2)\
, \label{J-reg}
\end{equation}
where
\begin{eqnarray}
J_{\pi^-}^{{\rm (mag)}}(k,\Pi^2) &=& \dfrac{N_c}{2\pi^2}
\int_0^\infty\! dz \int_{0}^1 dy \;\exp\big[-z M^2-z y(1-y)\Pi^2\big]
\,\times
\nonumber \\
&& \bigg\{ \bigg[ M^2 +\dfrac{1}{z} - y(1-y)\Big[\Pi^2-(2k+1)\,B_e\Big] \bigg] \times \nonumber \\
&& \bigg[\,\dfrac{\alpha_-^k}{\alpha_+^{k+1}}\,(1-t_u \,t_d)\, \exp\big[z\,y(1-y)(2k+1)B_e\big]
- \dfrac{1}{z} \bigg] +
\nonumber \\
&& \dfrac{\alpha_-^{k-1}}{\alpha_+^{k+2}}\,(1-t_u^2)\,(1-t_d^2)\,
\Big[ \alpha_- + (\alpha_- - \alpha_+)\,k \Big]
\,\times\nonumber \\
&& \exp\big[z\,y(1-y)(2k+1)\,B_e\big] - \dfrac{1}{z} \bigg[ \dfrac{1}{z} -
y(1-y)(2k+1)\,B_e \bigg] \bigg\}\ . \label{J-mag}
\end{eqnarray}
The integrand in Eq.~(\ref{J-mag}) is well behaved in the limit $z\to 0$.
Hence, this magnetic field-dependent contribution is finite. On the other
hand, the expression for the subtracted $B=0$ piece is the same as in the
$\pi^0$ case, Eq.~(\ref{TPFB0}), replacing $q^2\to \Pi^2$. Therefore, using
3D cutoff regularization, the function $J^{\rm (reg)}_{\pi,B=0}$ in
Eq.~(\ref{J-reg}) will be given by Eq.~(\ref{J+0reg}).

Given the regularized polarization function, we can now derive an equation
for the $\pi^-$ meson pole mass in the presence of the magnetic field. To do
this, let us first consider a pointlike pion. For such a particle, in
Euclidean space, the two-point function will vanish (i.e., the propagator
will have a pole) when
\begin{equation}
\Pi^2 = - m_{\pi^-}^2\ ,
\label{meB}
\end{equation}
or, equivalently, $q_\parallel^2 = - [ m_{\pi^-}^2 + (2k+1)\,eB]$, for a
given value of $k$. Therefore, in our framework the
charged pion pole mass can be obtained for each Landau level $k$ by solving the
equation
\begin{equation}
\frac{1}{2G} - J_{\pi^-}^{{\rm (reg)}}(k,-m_{\pi^-}^2) \ = \ 0 \ .
\end{equation}
While for a pointlike pion $m_{\pi^-}$ is a B-independent quantity (the
$\pi^-$ mass in vacuum), in the present model---which takes into account
the internal quark structure of the pion---this pole mass turns out to
depend on the magnetic field. Instead of dealing with this quantity, it has
become customary in the literature to define the $\pi^-$ ``magnetic
field-dependent mass'' as the lowest quantum-mechanically allowed energy of
the $\pi^-$ meson, namely
\begin{equation}
E_{\pi^-}(eB) \ = \  \sqrt{m_{\pi^-}^2 + (2k+1)\,eB + q_3^2}\;\Big|_{q_3 =0,\,k=0}
\ = \ \sqrt{m_{\pi^-}^2 + eB} \
\label{epimas}
\end{equation}
(see e.g.~Ref.~\cite{Bali:2017ian}). Notice that this ``mass'' is magnetic
field dependent even for a pointlike particle. In fact, owing to zero-point
motion in the 1-2 plane, even for $k=0$ the charged pion cannot be at rest
in the presence of the magnetic field.

\subsection{Pion field redefinition and quark-meson coupling constants}

As usual, the pion field wave function has to be redefined. In the
absence of an external magnetic field we have $\vec \pi(q)=Z_{\pi}^{1/2} \,
\tilde{\vec \pi}(q)$, where $Z_{\pi}$ is usually called the ``wave
function renormalization constant.'' It is defined by fixing the residue of
the two-point function at the pion pole. One has
\begin{equation}
Z_\pi^{-1} \,=\, g_{\pi qq}^{-2} \,=\, -\dfrac{\partial J_\pi(q^2)}{\partial q^2} \bigg\rvert_{q^2=-m_\pi^2}\ ,
\end{equation}
where $J_\pi(q^2)$ is the polarization function. Then, in the vicinity of
the pole, the action reads
\begin{equation}
S_{\pi}^{\,\mbox{\tiny quad}} \, \simeq \, \dfrac{1}{2} \int
\delta\tilde{\vec \pi}(-q)\, (q^2+m_\pi^2)\, \delta\tilde{\vec \pi}(q)\ .
\end{equation}
As expected, the energy dispersion relation is isotropic in this context.

We consider now the situation in which the external magnetic field is
present. For the neutral pion, as shown in Eq.~(\ref{J0B}), the polarization
function $J_{\pi^0}^{\rm (reg)}(q_\perp^2,q_\parallel^2)$ depends in a
different way on perpendicular and parallel components of $q$. We expand the
action in Eq.~\eqref{actionquadpi0p} around the pion pole ($q_\perp=0$,
$q_\parallel^2=-m_{\pi^0}^2$), factorize out the parallel derivative, and
redefine the pion field according to $\pi^0(q)=Z_{\parallel}^{1/2} \,
\tilde{\pi}^0(q)$. This leads to
\begin{equation}
S^{\,\mbox{\tiny quad}}_{\pi^0} \ \simeq \ \dfrac{1}{2} \int_{q}
\delta\tilde{\pi}^0(-q) \, \left[ u_{\pi^0}^2\, q_\perp^2 +
q_\parallel^2+m_{\pi^0}^2 \right] \delta\tilde{\pi}^0(q) \ ,
\label{actionquadpi0p_2}
\end{equation}
where we have defined
\begin{equation}
Z_{\parallel}^{-1} \,=\, -\dfrac{dJ_{\pi^0}^{\rm (reg)}}{dq_\parallel^2}
\Big\rvert_{\!\!{\tiny\begin{array}{l}
               q_\perp^2 \!= 0 \\
               q_\parallel^2 = -m_{\pi^0}^2
             \end{array}}} \!\equiv\, g_{\pi^0qq}^{-2} \ ,
\qquad
Z_{\perp}^{-1} \,=\, -\dfrac{dJ_{\pi^0}^{\rm
(reg)}}{dq_\perp^2}\Big\rvert_{\!\!{\tiny\begin{array}{l}
               q_\perp^2 \!= 0 \\
               q_\parallel^2 = -m_{\pi^0}^2
             \end{array}}},
\qquad
u_{\pi^0}^2 = \dfrac{Z_{\parallel}}{Z_{\perp}}\ .
\label{u_pi}
\end{equation}
Denoting $M_0(y)=[M^2-y(1-y)m_{\pi^0}^2]^{1/2}$ and
$M_0^\Lambda(y)=[\Lambda^2+M_0(y)^2]^{1/2}$, from
Eqs.~(\ref{J0B}-\ref{I23d}) we obtain
\begin{eqnarray}
Z_{\parallel}^{-1}\,\dfrac{4\pi^2}{N_c}  &=& - 2\int_0^1 dy \,
\left[ \dfrac{\Lambda}{M_0^\Lambda(y)} + \ln
\left( \dfrac{M_0(y)}{\Lambda+M_0^\Lambda(y)} \right) -
\dfrac{\Lambda^3\,y(1-y)\,m_{\pi^0}^2}{2\,M_0(y)^2\,M_0^\Lambda(y)^3} \right] - \nonumber\\
& & \sum_f \int_0^\infty dz \int_0^1 dy \, e^{-z M_0(y)^2}\,y(1-y)\;\times \nonumber \\
& & \Bigg\{ \left[M^2+y(1-y)m_{\pi^0}^2+\dfrac{2}{z} \right] \left( 1 - \dfrac{zB_f}{\tanh(zB_f)} \right)
+ \dfrac{1}{z} - \dfrac{zB_f^2}{\sinh^2(zB_f)} \Bigg\}
\label{Z_par}
\end{eqnarray}
and
\begin{eqnarray}
Z_{\perp}^{-1}\, \dfrac{4\pi^2}{N_c} &=& - 2\int_0^1 dy \,
\left[ \dfrac{\Lambda}{M_0^\Lambda(y)} + \ln
\left( \dfrac{M_0(y)}{\Lambda+M_0^\Lambda(y)} \right) -
\dfrac{\Lambda^3\,y(1-y)\,m_{\pi^0}^2}{2\,M_0(y)^2\,M_0^\Lambda(y)^3} \right] - \nonumber\\
&& \sum_f \int_0^\infty dz \int_0^1 dy \, e^{-z M_0(y)^2}
\Bigg\{ - \gamma_f(y,z) \left( \dfrac{1}{z\tanh(zB_f)} + \dfrac{2B_f}{\sinh^2(zB_f)}
\right) +
\nonumber \\
&& \left[M^2+y(1-y)m_{\pi^0}^2 \right] \left[ y(1-y) - \dfrac{\gamma_f(y,z)}{\tanh(zB_f)} \right]
+ \dfrac{3y(1-y)}{z}  \Bigg\}\ ,
\label{Z_per}
\end{eqnarray}
where $\gamma_f(y,z)$ was defined in Eq.~\eqref{gamma_f}. It is seen that,
owing to the pion internal structure, the energy
dispersion relation is anisotropic in the presence of an external magnetic
field. Namely, as already stated in Ref.~\cite{Fayazbakhsh:2013cha}, one has
\begin{equation}
E_{\pi^0}^2 = -q_4^2 = u_{\pi^0}^2\, q_\perp^2 + q_3^2+m_{\pi^0}^2\ .
\end{equation}
The direct comparison of our results for the renormalization constants with
those quoted in Ref.~\cite{Fayazbakhsh:2013cha} is not possible due to the
fact that different regularization procedures were followed in each case (we
use the MFIR scheme, while in Ref.~\cite{Fayazbakhsh:2013cha} an ultraviolet
cutoff is introduced). However, we have found some discrepancies between
both results when comparing the corresponding unregularized expressions.
We will come back to this point in Sec.~IV.

For charged pions, the momentum in the plane perpendicular to the
external magnetic field is quantized in Landau levels $k$. The energy
dispersion relation reads in this case
\begin{equation}
E_{\pi^-}^2 = -q_4^2 = (2k+1)B_e + q_3^2 + m_{\pi^-}^2\ .
\end{equation}
The redefined (negative) charged pion field is given by $\pi^-_{\bar
q}=Z_{\pi^-}^{1/2} \, \tilde{\pi}^-_{\bar q}$, where
\begin{equation}
Z_{\pi^-}^{-1} \ = \ -\dfrac{dJ_{\pi^-}^{\rm (reg)}(k,\Pi^2)}{d\Pi^2}\bigg\rvert_{\Pi^2=-m_{\pi^-}^2}
\ \equiv \ g_{\pi^-qq}^{-2}\, \ .
\end{equation}
Explicitly, denoting $M_-(y)=[M^2-y(1-y)m_{\pi^-}^2]^{1/2}$ and
$M_-^\Lambda(y)=[\Lambda^2+M_-(y)^2]^{1/2}$, from Eq.~(\ref{J-mag}) we find
\begin{eqnarray}
&& Z_{\pi^-}^{-1} \, \dfrac{2\pi^2}{N_c} =  - \int_0^1 dy \,
\left[ \dfrac{\Lambda}{M_-^\Lambda(y)} + \ln
\left( \dfrac{M_-(y)}{\Lambda+M_-^\Lambda(y)} \right) -
\dfrac{\Lambda^3\,y(1-y)\,m_{\pi^-}^2}{2\,M_-(y)^2\,M_-^\Lambda(y)^3} \right] +
\nonumber\\
& & \int_0^\infty\! dz
\int_{0}^1 dy \, e^{-z M_-(y)^2}\, z\,y(1-y)
\Bigg\{ \left[ M^2 + y(1-y)\left(m_{\pi^-}^2+(2k+1)B_e\right) +\dfrac{2}{z} \right] \times  \nonumber \\
& & \left[\,\dfrac{\alpha_-^k}{\alpha_+^{k+1}}\, (1-t_u \,t_d)\; e^{z\,y(1-y)(2k+1)B_e}  - \dfrac{1}{z} \right] -
\, \dfrac{1}{z} \left[ \dfrac{1}{z} - y(1-y)(2k+1)B_e \right] \nonumber\\
& & +\,\dfrac{\alpha_-^{k-1}}{\alpha_+^{k+2}}\,(1-t_u^2)\,(1-t_d^2)
\big[ \alpha_- + (\alpha_- - \alpha_+)\,k \big] \; e^{z\,y(1-y)(2k+1)B_e}
\Bigg\}\ .
\end{eqnarray}
The definitions of $t_u$, $t_d$ and $\alpha_\pm$ have been given above, see
text below Eq.~(\ref{J-B}).

\subsection{Pion-to-vacuum vector and axial vector amplitudes and weak decay constants}

In order to obtain pion-to-vacuum vector and axial vector amplitudes, we
have to ``gauge'' the effective action by introducing a set of vector and
axial vector gauge fields, $W^{V,a}_\mu(x)$ and $W^{A,a}_\mu(x)$,
respectively. This is done by performing the replacement
\begin{equation}
\gamma_\mu \partial_\mu \rightarrow \gamma_\mu \partial_\mu -
i\ \frac{\tau^a}{2} \,
\sum_{C=V,A} \Gamma^C_\mu\  W^{C,a}_\mu(x)\ ,
\end{equation}
where $\Gamma^V_\mu =\gamma_\mu$ and $\Gamma^A_\mu = \gamma_\mu \gamma_5$.
Once this extended gauged effective action is built, the corresponding
pion-to-vacuum amplitudes are obtained as the derivative of this action with
respect to $W^{C,a}_\mu(x)$ and the redefined meson fields, evaluated at
$W^{C,a}_\mu(x)=0$ (here $C=V,A$ and $a=1,2,3$). Therefore, the relevant
terms in the action are those linear in the pion and gauge fields. This
piece of the action can be written as
\begin{equation}
S_{\pi W} \,=\, \sum_{C=V,A} \ \sum_{\sigma=0,\pm} \, \int d^4x \, d^4x'\
W_\mu^{C,-\sigma}(x) \, F^{C,\sigma}_\mu(x,x') \, \delta \pi^\sigma(x')\ ,
\label{S_piW}
\end{equation}
where $W^{C,\pm}_\mu=(W^{C,1}_\mu \mp i\,W^{C,2}_\mu)/\sqrt{2}$,
$W^{C,0}_\mu \equiv W^{C,3}_\mu$, while the functions
$F^{C,\sigma}_\mu(x,x')$ are defined as
\begin{eqnarray}
F^{C,0}_\mu(x,x') &=& -\,\dfrac{iN_c}{2} \sum_f \trmin_D \left[\mathcal{S}^{\mbox{\tiny MF},f}_{x,x'}\,
\gamma_5 \, \mathcal{S}^{\mbox{\tiny MF},f}_{x',x} \Gamma^C_\mu \right]\ , \label{F0_mu} \\
F^{C,-}_\mu(x,x') &=& -\,iN_c \, \trmin_D \left[\mathcal{S}^{\mbox{\tiny MF},d}_{x,x'} \,
\gamma_5 \, \mathcal{S}^{\mbox{\tiny MF},u}_{x',x} \Gamma^C_\mu \right]\ , \label{F-_mu} \\
F^{C,+}_\mu(x,x') &=& -\,iN_c \, \trmin_D \left[\mathcal{S}^{\mbox{\tiny MF},u}_{x,x'} \, \gamma_5
\, \mathcal{S}^{\mbox{\tiny MF},d}_{x',x} \Gamma^C_\mu \right]\ .
\end{eqnarray}

\subsubsection{Neutral pion amplitudes and form factors}

As in the analysis of the $\pi^0$ mass, we expand the neutral pion field in
Eq.~\eqref{S_piW} in the Fourier basis. Then, pion-to-vacuum amplitudes read
\begin{eqnarray}
H_{\mu,C}^0(x,\vec q\,) & = & \frac{1}{2}\langle 0 | \bar\psi(x)\,
\Gamma^C_\mu\, \tau^3\, \psi(x) | \tilde \pi^0(\vec q\,) \rangle
\, =\, -\,\dfrac{\partial S_{\pi W}}{\partial\delta\tilde\pi^0(q)\,\partial W^{C,0}_\mu(x)} \nonumber \\
& = & -\,Z_{\parallel}^{1/2}\! \int d^4x' e^{iqx'}\, F_\mu^{C,0}(x,x')\ .
\label{Hneutro}
\end{eqnarray}
Using Eqs.~(\ref{sfx}) and~(\ref{F0_mu}),
and taking into account that in this case the Schwinger phases cancel out, after
integrating over $x'$ we get
\begin{eqnarray}
H_{\mu,C}^0(x,\vec q\,) &=& Z_{\parallel}^{1/2}\, \dfrac{iN_c}{2} \; e^{iqx}\, \sum_f \int_{p\,v} (2\pi)^4 \,
\delta^{(4)}(q-v) \, \trmin_D \left[\tilde{S}^f_{p^+} \,\gamma_5\, \tilde{S}^f_{p^-} \Gamma^C_\mu \right]\
,
\end{eqnarray}
where, as in previous subsections, we have defined $p^\pm = p \pm v/2$.

For convenience, we consider the linear combinations
\begin{eqnarray}
H_{\parallel,C}^{0,\,\epsilon}(x,\vec q\,) &=& H_{4,C}^0(x,\vec q\,) + \epsilon \,H_{3,C}^0(x,\vec q\,)\ ,
\nonumber\\
H_{\perp,C}^{0,\,\epsilon}(x,\vec q\,) &=& H_{1,C}^0(x,\vec q\,) + i\epsilon \,H_{2,C}^0(x,\vec q\,)\ ,
\label{comb}
\end{eqnarray}
where $\epsilon =\pm 1$. Using the relations in Appendixes A and B, after some calculation we obtain
\begin{eqnarray}
H_{\parallel,V}^{0,\,\epsilon}(x,\vec q\,) &=&
-\epsilon\,q_\parallel^{\,-\epsilon}\ e^{iqx} \sum_f s_f \int_0^\infty \!dz \int_0^1 \!dy \, \mathcal{F}^0(y,z)\ ,
\nonumber\\
H_{\perp,V}^{0,\,\epsilon}(x,\vec q\,) &=& 0\ ,
\label{H_0}
\end{eqnarray}
and
\begin{eqnarray}
H_{\parallel,A}^{0,\,\epsilon}(x,\vec q\,) &=&
- i\ q_\parallel^\epsilon \ e^{iqx} \sum_f \int_0^\infty \!dz \int_0^1 \!dy\, \mathcal{F}^0(y,z) \,
\coth(z\,B_f)\ ,
\nonumber\\
H_{\perp,A}^{0,\,\epsilon}(x,\vec q\,) &=&
- i\,q_\perp^\epsilon\, e^{iqx} \sum_f \int_0^\infty \!dz \int_0^1 \!dy \,
\mathcal{F}^0(y,z) \,\dfrac{\cosh[(2y-1)zB_f]}{\sinh(z\,B_f)}\ ,
\end{eqnarray}
where we have defined $q_\parallel^\epsilon = q_4 + \epsilon\, q_3$,
$q_\perp^\epsilon = q_1 + i\epsilon\, q_2$, and
\begin{equation}
\mathcal{F}^0(y,z) \,=\, Z_{\parallel}^{1/2} \, \dfrac{N_cM}{8\pi^2} \,B_f\,
e^{-z\left[M^2+y(1-y)q_\parallel^2 \right]} \, e^{-\gamma_f(y,z)\,q_\perp^2/B_f} \ .
\label{F^0}
\end{equation}
Now, following the notation of Ref.~\cite{Coppola:2018ygv}, we define the
neutral pion decay form factors by
\begin{eqnarray}
H_{\parallel,A}^{0,\,\epsilon}(x,\vec q\,) &=& - i \,q_\parallel^\epsilon \, e^{iqx} f_{\pi^0}^{(A1)} \ , \nonumber\\
H_{\perp,A}^{0,\,\epsilon}(x,\vec q\,) &=& - i \, q_\perp^\epsilon \, e^{i q x} \left[ f_{\pi^0}^{(A1)} - \epsilon\, f_{\pi^0}^{(A2)}
- f_{\pi^0}^{(A3)}\right] \ , \nonumber\\
H_{\parallel,V}^{0,\,\epsilon}(x,\vec q\,) &=& -\epsilon\,q_\parallel^{\,-\epsilon}\,e^{iqx}\, f_{\pi^0}^{(V)}
\label{fpi0_def}
\end{eqnarray}
(note that we are working in Euclidean space; therefore, the relations
$H_4=iH^0$ and $q_4=iq^0$ need to be considered when comparing with the
expressions in Ref.~\cite{Coppola:2018ygv}). In this way, for an on-shell
pion in its rest frame, i.e.~taking $q_\mu=im_\pi\delta_{\mu 4}$, the axial
decay constants are given by
\begin{eqnarray}
f_{\pi^0}^{(A1)} &=& Z_{\parallel}^{1/2} \dfrac{N_c M}{8\pi^2} \sum_f \int_0^\infty dz \int_{0}^1 dy
\; e^{-z M_0(y)^2} \dfrac{B_f}{\tanh(zB_f)}\ , \nonumber\\
f_{\pi^0}^{(A2)} &=& 0\ , \nonumber\\
f_{\pi^0}^{(A3)} &=& Z_{\parallel}^{1/2} \dfrac{N_c M}{8\pi^2} \sum_f\, 2\,B_f \int_0^\infty dz \int_{0}^1 dy
\; e^{-z M_0(y)^2}\, \gamma_f(y,z) \ ,
\label{axialneutral}
\end{eqnarray}
while the vector decay constant reads
\begin{eqnarray}
f_{\pi^0}^{(V)} &=& Z_{\parallel}^{1/2} \dfrac{N_c M}{8\pi^2} \sum_f \,
s_fB_f \int_0^\infty \!dz \int_{0}^1 dy \;
e^{-z M_0(y)^2} \ ,
\label{vectorneutral}
\end{eqnarray}
where $M_0(y)=[M^2-y(1-y)m_{\pi^0}^2]^{1/2}$ and $\gamma_f(y,z)$ is
defined in Eq.~\eqref{gamma_f}. It is seen that $f_{\pi^0}^{(A2)}$ vanishes,
as indicated from the general analysis in Ref.~\cite{Coppola:2018ygv}. Thus,
we find that in the presence of the external magnetic field there are in
general two axial and one vector nonvanishing form factors for the neutral
pion. Notice that in the chosen frame both $H_{\perp,V}^{0,\epsilon}$ and
$H_{\perp,A}^{0,\epsilon}$ are zero, hence $f_{\pi^0}^{(A3)}$ will not
contribute to the amplitudes.

It can be easily seen that $f_{\pi^0}^{(A3)}$ and $f_{\pi^0}^{(V)}$ are
finite and vanish in the $B\rightarrow 0$ limit. On the contrary, the
expression for $f_{\pi^0}^{(A1)}$ in Eq.~(\ref{axialneutral}) is divergent.
It can be regularized in the context of the MFIR scheme, i.e., subtracting
the corresponding divergent contribution in the $B = 0$ limit and adding
it in a regularized form, $f_{\pi^0,B=0}^{\rm (reg)}$. One has
\begin{equation}
f_{\pi^0}^{(A1),{\rm (reg)}} \ = \ f_{\pi^0,B=0}^{\rm (reg)} +  f_{\pi^0}^{(A1),{\rm (mag)}} \ ,
\end{equation}
where
\begin{equation}
f_{\pi^0}^{(A1),{\rm (mag)}} \ = \ Z_\parallel^{1/2}\, \dfrac{N_c M}{8\pi^2}
\sum_f B_f \int_0^\infty \!dz \int_{0}^1 \!dy \, e^{-z M_0(y)^2} \left[
\dfrac{1}{\tanh(zB_f)} - \dfrac{1}{zB_f} \right]\ .
\end{equation}
The divergent $B=0$ piece,
\begin{equation}
f_{\pi,B=0} \ = \ Z_{\pi}^{1/2}\, \dfrac{N_c M}{4\pi^2}
\int_0^\infty dz \int_{0}^1 \dfrac{dx}{z} \, e^{-zM_0(y)^2}\ ,
\label{fpi0B=0}
\end{equation}
can be regularized using a 3D momentum cutoff scheme, as done in the
previous subsections. One has in this way
\begin{equation}
f_{\pi^0,B=0}^{\rm (reg)} \ = \ -2\, Z_{\parallel}^{1/2} N_c M I_2(-m_{\pi^0}^2) \ ,
\label{fpi0B=0reg}
\end{equation}
where $I_2$ is given by Eq.~(\ref{I23d}). Note that we do not take the $B\to
0$ limit in $Z_\parallel$ (strictly, one should first regularize the form
factor and then redefine the pion wave function).

Finally, we find it convenient to define ``parallel'' and
``perpendicular'' axial decay constants $f_{\pi^0}^{(A\parallel)}$ and
$f_{\pi^0}^{(A\perp)}$, given in terms of $f_{\pi^0}^{(A1),{\rm (reg)}}$ and
$f_{\pi^0}^{(A3)}$ according to
\begin{equation}
f_{\pi^0}^{(A\parallel)} \ = \ f_{\pi^0}^{(A1),{\rm (reg)}}\ , \qquad\qquad
f_{\pi^0}^{(A\perp)} \ = \ f_{\pi^0}^{(A1),{\rm (reg)}} - f_{\pi^0}^{(A3)}\ .
\label{parallelandperp}
\end{equation}
Our expressions for the $\pi^0$ decay constants, taken before any
regularization scheme is applied, can be compared with those obtained in
Ref.~\cite{Fayazbakhsh:2013cha}. Although, as mentioned in the previous
subsection, we have found some discrepancies in the results for the
renormalization constants, it can be checked that the ratios
$f_{\pi^0}^{(A\parallel)} / g_{\pi^0qq}$ and
$f_{\pi^0}^{(A\perp)}/g_{\pi^0qq}$ are in agreement with those quoted in
Ref.~\cite{Fayazbakhsh:2013cha}, once different notations have been properly
compatibilized.

\subsubsection{Charged pion amplitudes and form factors}

As in the case of the polarization functions, we expand the charged pion
fields using Eq.~(\ref{Ritus}). Since the charged decay constants are real
and equal for both charged pions (we use the conventions in
Ref.~\cite{Coppola:2018ygv}), it is sufficient to consider the $\pi^-$
hadronic amplitudes
\begin{eqnarray}
H_{\mu,C}^-(x,\breve q) &=& \langle 0 | \bar\psi\, \Gamma^C_\mu\, \tau^+\,
\psi | \tilde \pi^-(\breve q) \rangle \, =\, -\, \sqrt{2}\, \dfrac{\partial
S_{\pi W}}{\partial\delta\tilde\pi^-_{\bar q}\,\partial W^{C,+}_\mu(x)} \nonumber\\
&=&
-\,\sqrt{2}\; Z_{\pi^-}^{1/2} \int_{x'} \mathbb{F}_{\bar q}^-(x')\,
F_\mu^{C,-}(x,x')\ ,
\label{Hcargado}
\end{eqnarray}
where $\bar{q}$ and $\breve{q}$ are defined as in Eq.~\eqref{notation2},
with $q_4 = iE_{\pi^-} = i\sqrt{m_{\pi^-}^2 + (2k+1)\,eB + q_3^2}$.
From Eqs.~(\ref{sfx}) and (\ref{F-_mu}) we have
\begin{equation}
H_{\mu,C}^-(x,\breve q)\, = \, i\sqrt{2}\,N_c\, Z_{\pi^-}^{1/2}\! \int \!d^4x'\;
\mathbb{F}_{\bar q}^-(x')\, e^{i[\Phi_d(x,x')+\Phi_u(x',x)]}
\! \int_{p\,v} \!\! e^{i v(x-x')}\, \trmin_D
\left[\tilde{S}_{p^+}^d\gamma_5\, \tilde{S}_{p^-}^u\,\Gamma^C_\mu \,
\right]\, .
\label{hmuc}
\end{equation}
For convenience, as in the $\pi^0$ case we concentrate on the linear
combinations $H_{\parallel,C}^{-,\,\epsilon}$ and
$H_{\perp,C}^{-,\,\epsilon}$, which are defined in a similar way as in
Eq.~(\ref{comb}). The expression in Eq.~(\ref{hmuc}) can be worked out
integrating first over $x'$. This leads to
\begin{eqnarray}
H_{\mu,C}^-(x,\breve q) & = & i\sqrt{2}\,N_c\, Z_{\pi^-}^{1/2}\,\frac{4\pi N_k}{B_e}
\;e^{iq_2 x_2}\,e^{iq_\parallel x_\parallel} \int_{p\,v_\perp} \!\!
\trmin_D \left[\tilde{S}_{p^+}^d\gamma_5\, \tilde{S}_{p^-}^u\,\Gamma^C_\mu \, \right]\Big|_{v_\parallel = q_\parallel}
\;\times \nonumber \\
& & e^{i v_1(x_1-x'_1)}\, D_k(\sqrt{2B_e}\,x'_1+\sqrt{2/B_e}\,q_2)\Big|_{x'_1=-x_1+2(v_2-q_2)/B_e}\ ,
\label{hmudk}
\end{eqnarray}
where for definiteness we have taken $B>0$. The relevant integrals over $p$
and $v_\perp$ can be calculated using the expressions for the traces quoted
in Appendix A and the relations in Appendix B. After some algebra one arrives at
\begin{eqnarray}
H_{\parallel,V}^{-,\,\epsilon}(x,\breve q) &=& -\epsilon\,\sqrt{2} \, q_\parallel^{\,-\epsilon}
\, \mathbb{F}^-_{\bar q}(x) \,  Z_{\pi^-}^{1/2}
\int_0^\infty \! dz \int_{0}^1 dy \ \mathcal{F}^-(z,y,q_\parallel^2)
\, (t_u - t_d) \ , \nonumber\\
H_{\perp,V}^{-,\,\epsilon}(x,\breve q) &=& 0 \ , \nonumber \\
H_{\parallel,A}^{-,\,\epsilon}(x,\breve q) &=& - i\,\sqrt{2}\, q_\parallel^{\,\epsilon}
\,\mathbb{F}_{\bar q}^-(x) \,  Z_{\pi^-}^{1/2}
\int_0^\infty \!dz \int_{0}^1 dy \ \mathcal{F}^-(z,y,q_\parallel^2)
\, (1-t_ut_d) \ , \nonumber\\
H_{\perp,A}^{-,\,\epsilon}(x,\breve q) &=& \epsilon\,\sqrt{2}\, \sqrt{B_e(2k+1+\epsilon)}
\,\mathbb{F}_{\bar{q}+\epsilon}^-(x) \,\times \nonumber \\
&& \, Z_{\pi^-}^{1/2} \int_0^\infty\! dz\int_0^1 dy \ \mathcal{F}^-(z,y,q_\parallel^2)
\, \left(\dfrac{\alpha_-}{\alpha_+}\right)^\epsilon \,(1+\epsilon\, t_u)(1+\epsilon\, t_d) \ ,
\label{haxial}
\end{eqnarray}
where
\begin{equation}
\mathcal{F}^-(z,y,q_\parallel^2) \ = \ \dfrac{N_c M}{4\pi^2}\;\dfrac{\alpha_-^k}{\alpha_+^{k+1}}
\; e^{-z[M^2+y(1-y)q_\parallel^2]}\ \ ,
\end{equation}
and $t_u$, $t_d$ and $\alpha_\pm$ are defined as in the text below Eq.~(\ref{J-B}). We have
also introduced the shorthand notation $\bar q+\epsilon=(k+\epsilon,q_2,q_3,q_4)$.

As in the case of the neutral pion, we follow the notation of
Ref.~\cite{Coppola:2018ygv}, defining the charged pion decay constants by
\begin{eqnarray}
H^{-,\,\epsilon}_{\parallel,V} (x,\breve q) & = &
-\epsilon \,\sqrt{2} \ f_{\pi^-}^{(V)} \ q_\parallel^{\,-\epsilon} \ \mathbb{F}^-_{\bar q}(x)
\ , \nonumber\\
H_{\parallel,A}^{-,\,\epsilon}(x,\breve q) &=& - i \,\sqrt{2} \, f_{\pi^-}^{(A1)} \ q_\parallel^{\,\epsilon} \
\mathbb{F}_{\bar q}^-(x)  \ , \nonumber\\
H_{\perp,A}^{-,\,\epsilon}(x,\breve q) &=& \epsilon \,\sqrt{2}\ \left[f_{\pi^-}^{(A1)} + \epsilon\,
f_{\pi^-}^{(A2)} - f_{\pi^-}^{(A3)} \right]\ \sqrt{B_e(2k+1+\epsilon)} \ \mathbb{F}_{\bar q+\epsilon}^-(x)
\label{chargedfpi}
\end{eqnarray}
where $q_\parallel^\epsilon=q_4+\epsilon\, q_3$. From
Eqs.~\eqref{haxial} and~\eqref{chargedfpi} we obtain
\begin{eqnarray}
f_{\pi^-}^{(A1)} &=& Z_{\pi^-}^{1/2} \int_0^\infty \!dz \int_{0}^1 dy \
\mathcal{F}^-(z,y,-E_{\pi^-}^2)\, (1-t_u\,t_d) \ , \nonumber\\
f_{\pi^-}^{(A2)} &=& Z_{\pi^-}^{1/2} \int_0^\infty \!dz \int_{0}^1 dy \
\mathcal{F}^-(z,y,-E_{\pi^-}^2)\, \left[ \frac{\alpha_-}{2 \alpha_+} (1+ t_u)(1+t_d)  - \frac{\alpha_+}{2 \alpha_-} (1- t_u)(1-t_d) \right]  \ , \nonumber\\
f_{\pi^-}^{(A3)} &=& Z_{\pi^-}^{1/2} \int_0^\infty \!dz \int_{0}^1 dy \
\mathcal{F}^-(z,y,-E_{\pi^-}^2)\,  \Big[1 - t_u\,t_d   \nonumber\\
& & \qquad \qquad\qquad \qquad\qquad \qquad \qquad \ \  - \frac{\alpha_-}{2 \alpha_+} (1+ t_u)(1+t_d)  - \frac{\alpha_+}{2 \alpha_-} (1- t_u)(1-t_d) \Big] \ , \nonumber\\
f_{\pi^-}^{(V)} &=& Z_{\pi^-}^{1/2} \int_0^\infty \!dz \int_{0}^1 dy \
\mathcal{F}^-(z,y,-E_{\pi^-}^2)\, (t_u-t_d) \ .
\label{fpis_charged}
\end{eqnarray}
Note that the
form factors have a dependence on $k$ and $B_e$ that has been omitted to
abbreviate the notation. In the $B\rightarrow 0$ limit we have $Z_{\pi^-}
\rightarrow Z_{\pi}$ and $f_{\pi^-}^{(A1)} \rightarrow f_{\pi,B=0}$, which
is given by Eq.~\eqref{fpi0B=0}. Meanwhile, $f_{\pi^-}^{(A2)}$,
$f_{\pi^-}^{(A3)}$ and $f_{\pi^-}^{(V)}$ are finite and vanish in the limit
$B\rightarrow 0$. Therefore, as expected, both neutral and charged pion weak
form factors tend to the usual pion decay constant in the absence of the
external field.

Once again, the expression for $f_{\pi^-}^{(A1)}$ in
Eq.~(\ref{fpis_charged}) is divergent and needs to be regularized. Using a
3D cutoff within the MFIR scheme, the regularized expression reads
\begin{equation}
f_{\pi^-}^{(A1),{\rm (reg)}} \ = \ f_{\pi^-,B=0}^{\rm (reg)} +
f_{\pi^-}^{(A1),{\rm (mag)}} \ ,
\end{equation}
where
\begin{equation}
f_{\pi^-}^{(A1),{\rm (mag)}} \ = \ Z_{\pi^-}^{1/2} \, \dfrac{N_c M}{4 \pi^2}
\int_0^\infty \!\!dz \int_{0}^1 \!dy \, e^{-z M_-(y)^2}
\left[\dfrac{\alpha_-^k}{\alpha_+^{k+1}}\,(1-t_u\,t_d)e^{zy(1-y)(2k+1)B_e} -\dfrac{1}{z} \right]\ ,
\end{equation}
with $M_-(y)=[M^2-y(1-y)m_{\pi^-}^2]^{1/2}$, and
\begin{equation}
f_{\pi^-,B=0}^{\rm (reg)} \ = \ -2 \, Z_{\pi^-}^{1/2}\, N_c\, M \,I_2(-m_{\pi^-}^2) \ ,
\label{fA1mag_charged}
\end{equation}
with $I_2(q^2)$ given by Eq.~\eqref{I23d}.

As in the case of the neutral pion, we find it convenient to introduce
parallel and perpendicular $\pi^-$ axial decay form factors. Thus, we define
one parallel and two perpendicular decay constants, according to
\begin{equation}
f_{\pi^-}^{(A\parallel)} \ = \ f_{\pi^-}^{(A1),{\rm (reg)}}\ , \qquad\qquad
f_{\pi^-}^{(A\perp\pm)} \ = \ f_{\pi^-}^{(A1),{\rm (reg)}} \pm f_{\pi^-}^{(A2)} -
f_{\pi^-}^{(A3)}\ .
\label{parallelandperpch}
\end{equation}
It is worth noticing that if the pion lies on the lowest Landau level,
i.e.~$k=0$, from Eq.~(\ref{chargedfpi}) one has
$H_{\perp,A}^{-,\,-}(x,\breve q) = 0$, hence in that case the $\pi^-$ weak
decay amplitude will not depend on $f_{\pi^-}^{(A\perp-)}$ [in fact,
strictly speaking, for $k=0$ one cannot determine $f_{\pi^-}^{(A\perp-)}$
from Eqs.~\eqref{haxial} and~\eqref{chargedfpi}].

The $\pi^+$ decay constants can be obtained following a similar procedure.
As stated in Ref.~\cite{Coppola:2018ygv}, one can check that
$f_{\pi^+}^{(i)}=f_{\pi^-}^{(i)}$, where $i=V,A1,A2,A3$. We recall that the
above expressions correspond to the case $B > 0$. By changing $B\rightarrow
-B$ one can see that
\begin{eqnarray}
f_{\pi^\pm}^{(V)}(k,B) & = & -f_{\pi^\pm}^{(V)}(k,-B) \ ,
\nonumber \\
f_{\pi^\pm}^{(Aj)}(k,B) & = & f_{\pi^\pm}^{(Aj)}(k,-B)\ ,\quad
j=1,2,3\ .
\end{eqnarray}

\section{Chiral limit relations}

It is interesting to discuss the relations satisfied by the quantities
studied in the previous section in the chiral limit, i.e., for
$m_0\to 0$. First, it should be stressed that even in the presence of an
external magnetic field, the neutral pion remains being a pseudo-Nambu-Goldstone (NG) boson. 
This can be shown by taking into account
the polarization function $J^{\rm (reg)}_{\pi^0}(q_\parallel^2,q_\perp^2)$
evaluated at $q_\parallel^2 = q_\perp^2 = 0$. After integration by parts it
is seen that $J^{\rm (mag)}_{\pi^0}(0,0)=2N_c\, I^{\rm (mag)}$,
where $I^{\rm (mag)}$ is given by Eq.~(\ref{imag}). Hence, from
Eqs.~(\ref{ireg}), (\ref{I13d}) and (\ref{J+0reg}) one gets
\begin{equation}
J^{\rm (reg)}_{\pi^0}(0,0) \ = \ 2N_c\, I^{\rm (reg)}\ .
\label{jregch}
\end{equation}
Now, taking into account this result together with the (regularized) gap
equation (\ref{gapeq}), in the chiral limit one gets $J^{\rm
(reg)}_{\pi^0}(0,0)_{\rm ch} = 1/(2G)$, which implies $m_{\pi^0,\,{\rm ch}}
=0$. In this way, associated chiral relations are expected to hold even for
nonzero $B$.

{}From the expressions for the renormalization constants,
Eqs.~(\ref{Z_par}-\ref{Z_per}), and the axial form factors,
Eq.~(\ref{axialneutral}), it is seen that the parallel and perpendicular
axial decay constants for the $\pi^0$ meson introduced in
Eq.~(\ref{parallelandperp}) satisfy the generalized Goldberger-Treiman
relations
\begin{eqnarray}
g_{\pi^0 qq}\, f_{\pi^0}^{(A\parallel)} & = & M_{\rm ch} + {\cal O}(m^2_{\pi^0}) \ ,
\label{gt1}\\
g_{\pi^0 qq}\, f_{\pi^0}^{(A\perp)} & = & u^2_{\pi^0,\,{\rm ch}}\, M_{\rm ch} + {\cal O}(m^2_{\pi^0})
\ .
\label{gt}
\end{eqnarray}
Thus, in the chiral limit one has
\begin{equation}
f_{\pi^0,\,{\rm ch}}^{(A\perp)}\ =\ u^2_{\pi^0,\,{\rm ch}}\;
f_{\pi^0,\,{\rm ch}}^{(A\parallel)}\ .
\label{Chiral}
\end{equation}
In fact, this equation can be readily obtained from a general effective low
energy action for NG bosons in the presence of a magnetic field, see
e.g.~Ref.~\cite{Miransky:2002rp}. Making use of Eq.~(\ref{gt1}), together
with the gap equation, one obtains the generalized Gell-Mann-Oakes-Renner
relation
\begin{equation}
\left( m_{\pi^0}^2\, f_{\pi^0,\,{\rm ch}}^{(A\parallel)} \right)^2 \ = \
-\,\frac{m_0}{2}\,
\langle\bar u u + \bar d d \rangle_{\rm ch} \ ,
\label{GMOR}
\end{equation}
where we have taken into account that in our model the averaged quark
condensate satisfies $\langle \bar uu + \bar dd \rangle/2
= - M_{\rm ch}/(2G) + {\cal O}(m_0)$. Note that a similar relation can be found
for $f_{\pi^0,\,{\rm ch}}^{(A\perp)}$ using Eq.~\eqref{Chiral}.

It is also interesting to consider the expression for $f_{\pi^0}^{(V)}$ in
the chiral limit. From Eqs.~(\ref{vectorneutral}) and (\ref{gt1}) it is seen
that for $m_0\to 0$ one has
\begin{eqnarray}
f_{\pi^0,\,{\rm ch}}^{(V)} \ = \ \frac{eB}{8 \pi^2 f_{\pi^0,\,{\rm ch}}^{(A\parallel)}}\ .
\label{chiralfv}
\end{eqnarray}
It is worth noticing that this result can be obtained from the anomalous
Wess-Zumino-Witten (WZW) effective Lagrangian~\cite{WZW}. The WZW term
that couples a neutral pion to an electromagnetic field and a vector field
$W_\mu^{V,3}$ is given by
\begin{equation}
{\cal L}_{WZW}\Big|_{\pi^0 A W^V} = \frac{i\,N_c\, e}{48 \pi^2 f_\pi}\;\pi^0
\; \epsilon_{\mu\nu\alpha\beta} \; \partial_\mu W_\nu^{V,3} \;
F_{\alpha\beta}\ ,
\label{lwzw}
\end{equation}
where $\epsilon_{4123}= 1$. If one identifies the constant $f_\pi$ in this
effective Lagrangian with $f_{\pi^0}^{(A\parallel)}$, and the electromagnetic
field tensor with the external magnetic field ($F_{12}=-F_{21}=B$), taking
into account the definitions in Eq.~(\ref{fpi0_def}) one arrives at the
chiral relation in Eq.~(\ref{chiralfv}).

In the case of charged pions, the presence of an external magnetic
field leads to the explicit breakdown of chiral symmetry and, in general,
$\pi^\pm$ cannot be identified with NG bosons. However, chiral relations
should be recovered in the limit of low $eB$. In particular, the coupling
of charged pions to the magnetic field and an external vector current
arising from the WZW Lagrangian has the same form of Eq.~(\ref{lwzw}),
taking the $i=1,2$ isospin components of the fields $\pi^i$ and
$W_\nu^{V,i}$.

\section{Numerical results}

To obtain some numerical results for the different pion properties one has
to fix the model parametrization. Here, as done in
Ref.~\cite{Coppola:2018vkw}, we take the parameter set $m_0 = 5.66$ MeV,
$\Lambda = 613.4$ MeV and $G\Lambda^2 = 2.250$, which (for vanishing
external field) corresponds to an effective mass $M=350$~MeV and a
quark-antiquark condensate $\langle \bar f f\rangle_0 = (-243.3\ {\rm
MeV})^3$. This parametrization, denoted as set I, properly reproduces
the empirical values of the pion mass and decay constant in
vacuum, namely $m_\pi=138$~MeV and $f_\pi=92.4$~MeV. It also provides a very
good agreement with the results from lattice QCD quoted in
Ref.~\cite{Bali:2011qj} for the normalized average condensate $\Delta
\bar\Sigma(B)$~\cite{Coppola:2018vkw}. To test the sensitivity of our
results to the model parametrization we have also considered two
alternative parameter sets, denoted as set II and set III, which also
reproduce the phenomenological values of $m_\pi$ and $f_\pi$ in vacuum, and
lead to effective masses $M=320$ and 380~MeV, respectively.

\subsection{Neutral pion}

\begin{figure}[htb]
\centering{\includegraphics[width=1.\textwidth]{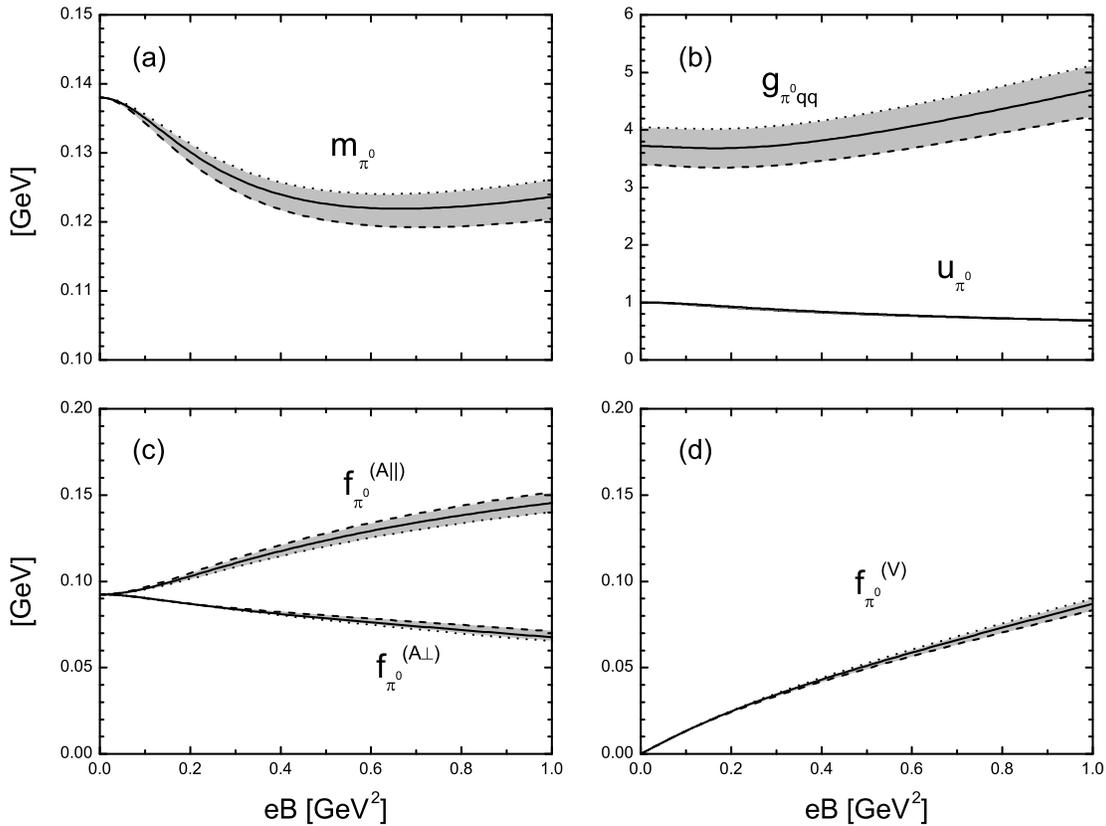}}
\vspace{-1.5cm}
\caption{Neutral pion properties as functions of $eB$. Solid lines
correspond to set I, while the limits of the gray bands correspond to
set II (dashed lines) and set III (dotted lines).}
\label{Fig_pi0}
\end{figure}

In Fig.~\ref{Fig_pi0} we show our numerical results for the quantities
associated with the neutral pion as functions of $eB$. Solid lines
correspond to the results from set I, while the limits of the grey band
correspond to those from set II (dashed lines) and set III (dotted lines).
We observe that the qualitative behavior of all calculated quantities
remains basically unaffected by changes in the model parameters within
phenomenologically reasonable limits. The results for the pion mass, shown
in Fig.~\ref{Fig_pi0}(a), have already been given in
Ref.~\cite{Coppola:2018vkw}, and are included here just for completeness. It
is seen that the mass shows a slight decrease with $eB$, which is also in
agreement with the analysis in
Refs.~\cite{Avancini:2015ady,Avancini:2016fgq}. Some lattice simulations
using Wilson fermions~\cite{Bali:2017ian} seem to favor a somewhat larger
decrease of $m_{\pi^0}$ as the magnetic field increases. In these simulations,
however, a heavy pion with mass $m_\pi(0)=415$~MeV in vacuum has been
considered. It is interesting to note that in the framework of NJL-like
models some enhancement of the decrease can be obtained either by assuming a
magnetic field dependent coupling constant~\cite{Avancini:2016fgq} or by
considering nonlocal interactions~\cite{GomezDumm:2017jij}.

In Fig.~\ref{Fig_pi0}(b) we plot the coupling constant $g_{\pi^0qq}$ and the
directional refraction index $u_{\pi^0}$, given by Eqs.~\eqref{u_pi}
and~\eqref{Z_par}. We observe that $g_{\pi^0qq}$ shows some enhancement if
$B$ is increased. On the other hand, $u_{\pi^0}$ decreases monotonously with
$eB$, remaining always lower than one. These results are
consistent with those obtained in
Refs.~\cite{Gusynin:1995nb,Fukushima:2012kc}. It should be also noticed that
$u_{\pi^0}$ is basically insensitive to the parametrization. In fact, it is
kept almost unchanged if one takes $m_0\rightarrow 0$, which implies that
for nonzero $B$ neutral pions move at a speed lower than the speed of light
even in the chiral limit. We notice that, on the contrary, in
Ref.~\cite{Fayazbakhsh:2013cha} it is found that $u_{\pi^0} > 1$. It is
unclear to us whether this different behavior is due to the already
mentioned discrepancies in the expressions for the renormalization constants
or to the different procedures chosen for the regularization.

Our results for the neutral axial decay constants are shown in
Fig.~\ref{Fig_pi0}(c). Starting from a common value at $B=0$, it is seen
that while $f_{\pi^0}^{(A\parallel)}$ gets enhanced for increasing $eB$,
$f_{\pi^0}^{(A\perp)}$ gets reduced. In both cases the $B$ dependence is
stronger than for the other quantities discussed previously. Note that our
results indicate that $f_{\pi^0}^{(A\perp)} < f_{\pi^0}^{(A\parallel)}$ for
all considered values of $eB$, which differs from the result in
Ref.~\cite{Fayazbakhsh:2013cha}. This seems to be related to the fact that,
as stated, in that paper $u_{\pi^0} > 1$ is obtained. Finally, in
Fig.~\ref{Fig_pi0}(d) we show the behavior of $f_{\pi^0}^{(V)}$ as a
function of $eB$. It is seen that, starting from 0 at $eB=0$, the vector
decay constant grows with $eB$, reaching a value comparable to the
average of the axial decay constants $f_{\pi^0}^{(A\parallel)}$ and
$f_{\pi^0}^{(A\perp)}$ at $eB \sim 1$~GeV$^2$.

It is interesting to notice that the numerical results given above (which
have been obtained from parametrization sets leading to $m_{\pi^0}=138$~MeV
at $B=0$) satisfy quite well the chiral limit relations in
Eqs.~(\ref{gt1}-\ref{chiralfv}). In fact, it is found that all these
relations are satisfied at a level of less than 1\% for all considered
values of $eB$.

\begin{figure}[htb]
\centering{\includegraphics[width=1.\textwidth]{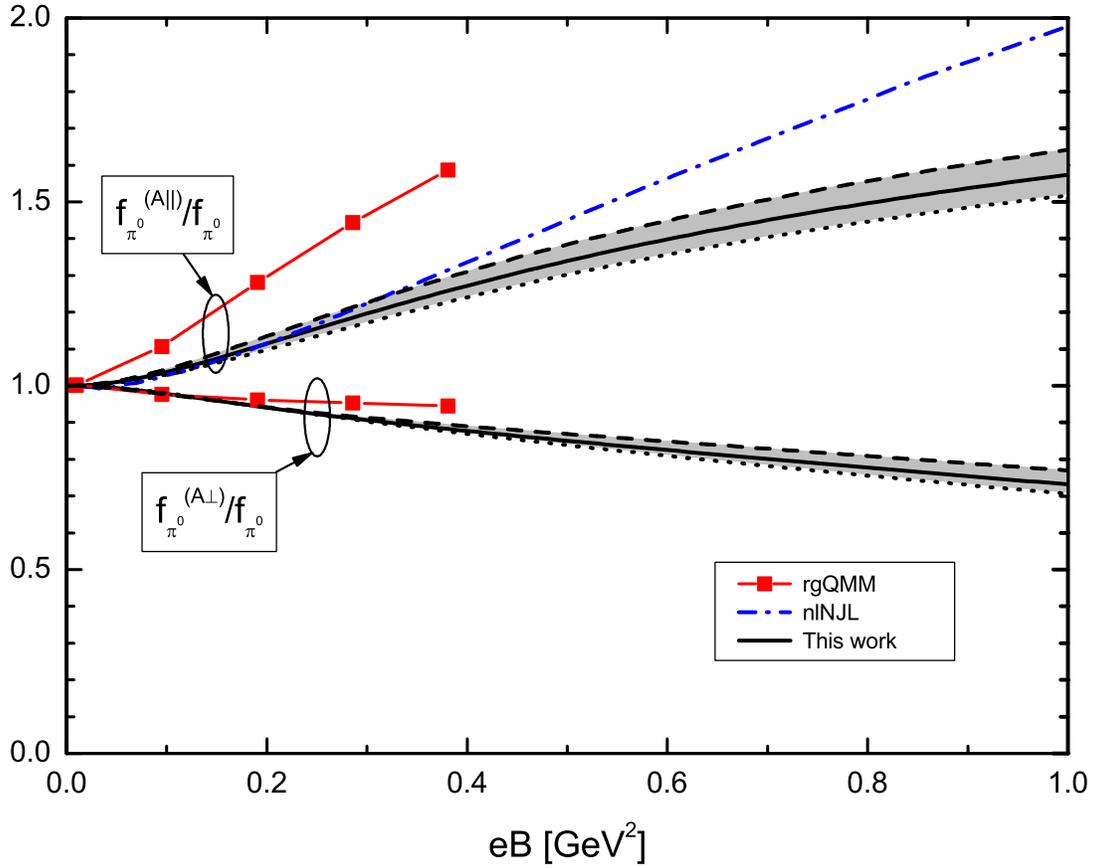}}
\vspace{-1.5cm}
\caption{Neutral pion decay constants as functions of $eB$ for various models.}
\label{Fig_comp}
\end{figure}

To conclude this subsection, in Fig.~\ref{Fig_comp} we show a comparison
between our results for the axial decay constants, normalized to the value
at $B=0$, and the results obtained in Refs.~\cite{GomezDumm:2017jij}
and~\cite{Kamikado:2013pya}. Those works are based on a nonlocal NJL model
(nlNJL), dashed-dotted line in the figure, and on the functional renormalization
group approach to the quark-meson model (rgQMM), red squares,
respectively. We see that in the case of $f_{\pi^0}^{(A\parallel)}$ our
results are somewhat below those obtained within the rgQMM. This is likely
to be correlated with the fact that in that approach the $\pi^0$ mass shows
a stronger decrease as the magnetic field increases. A similar trend is
found for $f_{\pi^0}^{(A\perp)}$, although in this case the difference with
the rgQMM calculation of Ref.~\cite{Kamikado:2013pya} is somewhat smaller.
It should be mentioned that additional calculations for
$f_{\pi^0}^{(A\parallel)}$ have been carried out using
ChPT~\cite{Andersen:2012zc} and within the effective chiral confinement
Lagrangian approach~\cite{Andreichikov:2018wrc}. The latter shows a behavior
similar to that of the nlNJL model considered in
Ref.~\cite{GomezDumm:2017jij}, while ChPT results, trustable for values of
the magnetic field up to say $eB \sim 0.1$~GeV$^2$, are found to be in
reasonable agreement with our curves.

\subsection{Charged pions}

\begin{figure}[htb]
\centering{\includegraphics[width=1.\textwidth]{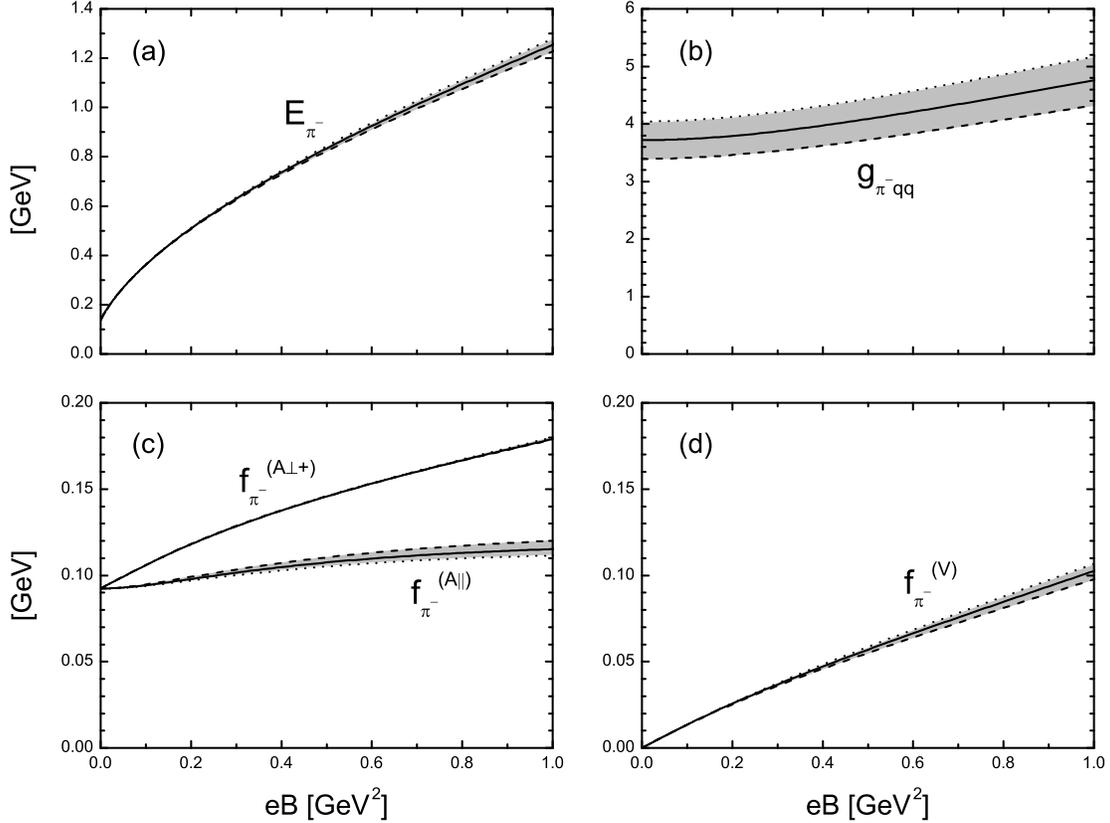}}
\vspace{-1.5cm}
\caption{Charged pion decay properties as functions of $eB$. Solid lines
correspond to set I, while the limits of the grey bands correspond to
set II (dashed lines) and set III (dotted lines).}
\label{Fig_pim}
\end{figure}

In Fig.~\ref{Fig_pim} we show our numerical results for the quantities
associated with charged pions, in the lowest Landau level (LLL), as
functions of $eB$. As in the previous subsection, solid lines indicate the
results for parameter set I, while the limits of the gray bands correspond
to set II (dashed lines) and set III (dotted lines). From the figure it is
observed that, as in the case of the $\pi^0$, the qualitative behavior of
all calculated quantities is not significantly affected by changes in the
model parametrization within the considered limits. In Fig.~\ref{Fig_pim}(a)
we quote the results for the magnetic field-dependent charged pion mass, see
Eq.~(\ref{epimas}), which have already been presented in
Ref.~\cite{Coppola:2018vkw}. They are included here just for completeness.
As discussed in Ref.~\cite{Coppola:2018vkw}, our results are in fair
agreement with those obtained from LQCD~\cite{Bali:2017ian}, once the
current quark mass is increased so that $m_{\pi^+}(B=0)$ matches the value
of the pion mass considered in lattice calculations.
In Fig.~\ref{Fig_pim}(b) we quote the curves corresponding to the coupling
constant $g_{\pi^-qq}$ as a function of $eB$. It can be seen that they are
quite similar to those obtained for the neutral pion in Fig.~\ref{Fig_pi0}(b). The behavior of the
axial decay constants is shown in Fig.~\ref{Fig_pim}(c). We choose to plot
$f_{\pi^-}^{(A\parallel)}$ (also denoted as $f_{\pi^-}^{(A1),({\rm reg})}$) and the
combination $f_{\pi^-}^{(A\perp+)} = f_{\pi^-}^{(A1),({\rm reg})} + f_{\pi^-}^{(A2)} -
f_{\pi^-}^{(A3)}$, since ---as discussed in Sec.~II--- if the pion
lies on the LLL these are the only relevant form factors for the evaluation
of the matrix elements of the axial current. From the figure it is observed
that $f_{\pi^-}^{(A\parallel)}$ shows a slight growth with $eB$, lower than
that of $f_{\pi^0}^{(A\parallel)}$ [see Fig.~\ref{Fig_pi0}(c)]. On the other hand,
$f_{\pi^-}^{(A\perp+)}$ exhibits a strong increase with $eB$, reaching a
magnitude of about 180~MeV for $eB=1$~GeV$^2$. Finally, in Fig.~\ref{Fig_pim}(d)
we plot $f_{\pi^-}^{(V)}$ as a function of the magnetic field. Its behavior
is similar to that of $f_{\pi^0}^{(V)}$, shown in Fig.~\ref{Fig_pi0}(d).

In the framework of lattice QCD, some results for $f_{\pi^-}^{(A\parallel)}$
and $f_{\pi^-}^{(V)}$ in the presence of an external magnetic field have
been presented recently~\cite{Bali:2018sey}. Although errors are still
relatively large, it can be seen that beyond the first lattice data points
$f_{\pi^-}^{(A\parallel)}$ shows an overall increase with the magnetic
field, in qualitative agreement with our results. On the other hand, a
continuum extrapolation seems to indicate that $f_{\pi^-}^{(A||)}$ starts
out with a negative slope, which differs from the case of
$f_{\pi^0}^{(A\parallel)}$. We find this result difficult to understand,
since the decay constants of charged and neutral pions should behave
similarly~\cite{Andersen:2012zc} for very small values of $eB$. In
addition, in Ref.~\cite{Dominguez:2018njv} the magnetic field dependence of
$f_{\pi^-}^{(A\parallel)}$ has been analyzed in the context of QCD sum
rules. In comparison with our results, their analysis shows a steeper
enhancement with $B$, leading to $f_{\pi^-}^{(A\parallel)}\sim 0.17$~GeV for
$eB= 1$~GeV$^2$. In any case, it should be stressed that our results show
that, as expected, the Goldberger-Treiman and Gell-Mann-Oakes-Renner
relations for charged pions [i.e., the equivalent to Eqs.~\eqref{gt1}
and~\eqref{GMOR}, obtained for neutral mesons] are violated for $eB \gtrsim
m_\pi^2$, for both $f_{\pi^-}^{(A\parallel)}$ and $f_{\pi^-}^{(A\perp+)}$.

To conclude, let us make an additional comment on the magnetic field
dependences of the decay constants. In the chiral limit, it can be seen that
for low values of $eB$ the difference $f_{\pi^-}^{(A2)} - f_{\pi^-}^{(A3)}$
is given by
\begin{equation}
f_{\pi^-,\,{\rm ch}}^{(A2)} - f_{\pi^-,\,{\rm ch}}^{(A3)} \ = \
\frac{eB}{8\pi^2f_{\pi^-,\,{\rm ch}}^{(A\parallel)}} \,
\Big( 1\, - \, \frac{7\,eB}{45\,M_{\rm ch}^2} \, +\, \dots\Big) \ .
\end{equation}
On the other hand, in the case of $f_{\pi^-}^{(V)}$, for low values of the
magnetic field a relation similar to Eq.~(\ref{chiralfv}) is expected to be
satisfied in the chiral limit. From our numerical calculations we find quite
remarkable that relations of the same form, i.e.,
\begin{equation}
f_{\pi^-}^{(A2)} - f_{\pi^-}^{(A3)} \ = \
\frac{eB}{8\pi^2f_{\pi^-}^{(A\parallel)}} \,
\Big( 1\, - \, \frac{7\,eB}{45\,M^2} \Big)
\label{fa2mas}
\end{equation}
and
\begin{eqnarray}
f_{\pi^-}^{(V)} \ = \ \frac{eB}{8 \pi^2 f_{\pi^-}^{(A\parallel)}}\ ,
\label{fvchir}
\end{eqnarray}
are in fact approximately valid also for large external magnetic fields.
Indeed, although in the presence of the magnetic field the $\pi^-$ cannot be
considered a pseudo-Goldstone boson, we find that $f_{\pi^-}^{(A2)} -
f_{\pi^-}^{(A3)}$ and $f_{\pi^-}^{(V)}$ can be approximated by the
expressions in Eqs.~(\ref{fa2mas}) and (\ref{fvchir}) within 15\% and 10\%
accuracy, respectively, for values of $eB$ up to 1~GeV$^2$. It would be
interesting to verify if equivalent relations also arise within other
theoretical approaches to low energy hadron physics.

\section{Summary and conclusions}

In this work we have considered the approach introduced in
Ref.~\cite{Coppola:2018vkw} for the study of pion masses, extending the
calculations to other properties of neutral and charged pions. Such an
approach is based on the usage of the Nambu-Jona-Lasinio effective model for
low energy QCD dynamics, in which pions are treated as quantum fluctuations
in the random phase approximation. While for the $\pi^0$ one can take the
usual momentum basis to diagonalize the corresponding polarization
functions, this is not possible in the case of charged pions, due to the
presence of nonvanishing contributions from Schwinger phases. Therefore, to
diagonalize the charged pion polarization function we use a method based on
the Ritus eigenfunction approach to magnetized relativistic systems. Since
the NJL model is not renormalizable, the calculation of observables requires
an appropriate regularization scheme in order to deal with ultraviolet
divergences. Here, we have used the magnetic field independent
regularization procedure, in which only divergent vacuum
contributions to quantities at zero external magnetic field are regularized.
This scheme has been shown to provide more reliable predictions in
comparison with other regularization methods often used in the
literature~\cite{Avancini:2019wed}. Within the framework just described, we
have concentrated in particular on the analysis of the quark-meson coupling
constants, the neutral pion directional refraction index $u_{\pi^0}$, and
the form factors associated with pion-to-vacuum matrix elements of the
vector and axial vector hadronic currents.

In the case of the neutral pion we find that while the coupling constant
$g_{\pi^0qq}$ shows some enhancement if the external magnetic field is
increased, $u_{\pi^0}$ decreases monotonously with $eB$, remaining always
lower than one. We have checked that $u_{\pi^0}$ is kept almost unchanged if
one takes $m_0\rightarrow 0$, which implies that, contrary to the result
obtained in Ref.~\cite{Fayazbakhsh:2013cha}, for nonzero $B$ neutral pions
move at a speed lower than the speed of light even in the chiral limit.
Concerning the study of pion-to-vacuum amplitudes, in agreement with
previous
analyses~\cite{Miransky:2002rp,Kamikado:2013pya,Fayazbakhsh:2012vr,Coppola:2018ygv}
we find that for the $\pi^0$, in the presence of the external magnetic
field, there are in general two axial nonvanishing form factors, namely
$f_{\pi^0}^{(A\parallel)}$ and $f_{\pi^0}^{(A\perp)}$. Moreover, as
discussed in Ref.~\cite{Coppola:2018ygv}, the vector hadronic current is
also found to be nonvanishing, and an additional vector form factor
$f_{\pi^0}^{(V)}$ can be defined. We have verified that in the chiral limit
these quantities satisfy some relations. In fact, apart from the well-known
generalized Goldberger-Treiman and Gell-Mann-Oakes-Renner equations for
$f_{\pi^0}^{(A\parallel)}$ (see e.g.~Ref.~\cite{Agasian:2001ym}), we show
that in that limit the relations $f_{\pi^0}^{(A\perp)}\ =\ u^2_{\pi^0}
f_{\pi^0}^{(A\parallel)}$ and $f_{\pi^0}^{(V)} \ = \ eB/(8 \pi^2
f_{\pi^0}^{(A\parallel)})$ hold. The first of these equations follows from
the expressions quoted in Ref.~\cite{Fayazbakhsh:2013cha} and can also be
derived in the context of ChPT. On the other hand, the second one can be
related to the anomalous Wess-Zumino-Witten effective lagrangian, and---to
the best of our knowledge---has not been previously stated in the
literature. Our numerical results for the neutral axial decay constants
indicate that, starting from a common value at $B=0$,
$f_{\pi^0}^{(A\parallel)}$ gets enhanced for increasing $eB$, while
$f_{\pi^0}^{(A\perp)}$ gets reduced. We see that in the case of
$f_{\pi^0}^{(A\parallel)}$ our results are somewhat below those obtained in
Refs.~\cite{Kamikado:2013pya,GomezDumm:2017jij}. This is likely to be
correlated with the fact that in those approaches the $\pi^0$ mass shows a
stronger decrease as the magnetic field increases. A similar trend is found
for $f_{\pi^0}^{(A\perp)}$, although in this case the difference with the
calculation of Ref.~\cite{Kamikado:2013pya} is somewhat smaller. It is
interesting to notice that the numerical results for the form factors,
obtained for model parameters leading to a physical $B=0$ pion mass, satisfy
chiral limit relations in Eqs.~(\ref{gt1}-\ref{chiralfv}) quite well (that
is, within 1\% for all considered values of $eB$).

For the charged pions we find that the $B$ dependence of the corresponding
quark-meson coupling constant is quite similar to the one found in the case of
the $\pi^0$. Concerning the axial form factors, we see that while in general
three decay constants can be defined~\cite{GomezDumm:2017jij}, only two
linear combinations of them, $f_{\pi^-}^{(A\parallel)}$ and
$f_{\pi^-}^{(A\perp+)}$, are physically relevant for charged pions in their
lowest energy state. As in the case of the $\pi^0$, we find that there is
also a vector form factor $f_{\pi^-}^{(V)}$ that can be
nonvanishing~\cite{Bali:2018sey,GomezDumm:2017jij}. Our numerical results
indicate that while $f_{\pi^-}^{(A\parallel)}$ shows a rather slight
growth with the magnetic field (somewhat lower than that of
$f_{\pi^0}^{(A\parallel)}$), $f_{\pi^-}^{(A\perp+)}$ exhibits a
stronger increase with $eB$, reaching a magnitude of about 180~MeV for
$eB=1$~GeV$^2$.
Finally, it is seen that for $eB\lesssim 1$~GeV$^2$ the decay
constants for the charged pion satisfy approximate relations that are
equivalent to those obtained in the chiral limit for low values of the
magnetic field.

\begin{acknowledgments}

This work has been supported in part by Consejo Nacional de Investigaciones
Cient\'ificas y T\'ecnicas and Agencia Nacional de Promoci\'on Cient\'ifica
y Tecnol\'ogica (Argentina), under Grants No.~PIP17-700 and No.
PICT17-03-0571, respectively ; by the National University of La Plata
(Argentina), Project No.~X824; by the Ministerio de Econom\'ia y
Competitividad (Spain), under Contract No.~FPA2016-77177-C2-1-P; and by the
Centro de Excelencia Severo Ochoa Programme, Grant No.~SEV-2014-0398.

\end{acknowledgments}

\section*{APPENDIX A: DIRAC TRACES}

In this appendix we provide the explicit form of the Dirac traces that
appear in the calculation of the pion two-point functions and the
pion-to-vacuum matrix elements. In all cases we use the Schwinger form of
the propagators, with $\tilde S_p^f$ given by Eq.~(\ref{sfp_schw}). As in
the main text, we separate the four-vectors into parallel and perpendicular
two-vectors, e.g.~$p_\perp = (p_1,p_2)$ , $p_\parallel = (p_3,p_4)$.

The traces appearing in the two-point functions can be written as
\begin{equation}
\trmin_D \left[ \tilde S_{p^+}^{f_1} \gamma_5\, \tilde S_{p^-}^{f_2}
\gamma_5 \right] \ = \ \int_0^\infty d\tau \int_0^\infty d\tau' \ \exp
\left[-\tau\phi_{f_1}(\tau,p^+)-\tau'\phi_{f_2}(\tau',p^-)\right] \, T_5\ ,
\label{line1}
\end{equation}
where
\begin{equation}
\phi_{f_i}(\tau,p^\pm) \ = \ M^2 \, + \, {p_\parallel^\pm\,}^2 \,
+\,\frac{t_{f_i}}{\tau B_f} \;{p_\perp^\pm\,}^2\ ,
\end{equation}
with $t_{f_i} = \tanh (\tau B_{f_i})$. Writing $p^\pm = p\pm v/2$, from
Eq.~(\ref{sfp_schw}) one has
\begin{equation}
T_5 \ = \ 4\,  \left[\left( M^2+ p_\parallel^2 - \frac{v_\parallel^2}{4}\,\right)\,
(1+s_{f_1}\,s_{f_2}\,t_{f_1} \,t'_{f_2}) \, + \, (1-t^2_{f_1})\, (1- {t'_{f_2}}^2)\,
\left(p_\perp^2-\frac{v_\perp^2}{4}\right)\right]\ ,
\end{equation}
where $t'_{f_i} = \tanh (\tau' B_{f_i})$. Similarly, for the traces appearing
in the analysis of the pion-to-vacuum matrix elements we write
\begin{equation}
\trmin_D \left[\tilde S_{p^+}^{f_1} \gamma_5 \, \tilde S_{p^-}^{f_2}
\Gamma^C_\mu \right] \ = \ \int_0^\infty d\tau \int_0^\infty d\tau' \ \exp
\left[-\tau\phi_{f_1}(\tau,p^+)-\tau'\phi_{f_2}(\tau',p^-)\right] \, T^C_\mu \ .
\label{MEpi0_1}
\end{equation}
Taking into account the linear combinations relevant for our calculations,
we find
\begin{eqnarray}
T_\parallel^{V,\epsilon} & = & T^V_4+\epsilon\, T^V_3 \, = \,
-4iM\, (v_3-\epsilon\,v_4)\,\big(s_{f_1}\,t_{f_1}+s_{f_2}\,t'_{f_2}\big)\ , \\
T_\parallel^{A,\epsilon} & = & T^A_4+\epsilon\, T^A_3 \, = \, -4M\, (v_4+\epsilon\,v_3)\,
\big(1\, + \, s_{f_1}\,s_{f_2}\,t_{f_1}\,t'_{f_2}\big)\ , \\
T_\perp^{V,\epsilon} & = & T^V_1+\epsilon\,i\, T^V_2 \, = \, 0 \ , \\
T_\perp^{A,\epsilon} & = & T^A_1+\epsilon\,i\, T^A_2 \, = \, -4M \bigg\{\! (p_1+\epsilon\, i p_2)
\Big[(1-t_{f_1}^2)\,(1+\epsilon\,s_{f_2}\, t'_{f_2}) -
(1-t_{f_2}^2)\,(1-\epsilon\,s_{f_1}\, t_{f_1})\Big] + \nonumber \\
& & \frac{1}{2}\,(v_1+\epsilon\, i v_2)\Big[(1-t_{f_1}^2)\,(1+\epsilon\,s_{f_2}\, t'_{f_2})
+ (1-{t'_{f_2}}^2)\,(1-\epsilon\,s_{f_1}\, t_{f_1})\Big] \bigg\} \ .
\end{eqnarray}

\section*{APPENDIX B: INTEGRALS OVER INTERNAL MOMENTA}

The integrals in Eqs.~(\ref{jota+}) and (\ref{hmudk}) can be performed using
the properties of the cylindrical parabolic functions $D_k(x)$. We need to
calculate
\begin{equation}
I_\lambda \ = \ \int_{p\,v_\perp}
e^{-\tau\phi_{f_1}(\tau,p^+)}\, e^{-\tau'\phi_{f_2}(\tau',p^-)} \,
e^{iv_1(x_1-x'_1)}\, D_k(\sqrt{2B_e}\,x'_1+\sqrt{2/B_e}\,q'_2)\; T_\lambda \
,
\label{h1}
\end{equation}
where $T_\lambda$ stands for the functions $T_5$,
$T_{\parallel}^{C,\epsilon}$ and $T_{\perp}^{C,\epsilon}$ in Appendix A.

The integrals over $p_\parallel$ can be easily obtained from the
relations
\begin{eqnarray}
\int_{p_\parallel} e^{-a (p_\parallel+v_\parallel/2)^2}\, e^{-b
(p_\parallel-v_\parallel/2)^2} & = & \frac{1}{4\pi(a+b)}\; e^{-\frac{a
b}{a+b}\;v_\parallel^2}\ ,
\nonumber \\
\int_{p_\parallel}
\left(p_\parallel^2-\frac{v_\parallel^2}{4}\right)\;
e^{-a (p_\parallel+v_\parallel/2)^2}\,e^{-b (p_\parallel-v_\parallel/2)^2} & = &
\frac{1}{4\pi(a+b)^2}\,\Big[ 1 - \frac{a b}{a+b}\; v_\parallel^2\Big]
\, e^{-\frac{a b}{a+b}\;v_\parallel^2}\ .
\end{eqnarray}
These expressions can be also applied for the integrals over $p_\perp$. For the case of
$T_\perp^{A,\epsilon}$ we also need
\begin{equation}
\int_{p_\perp}(p_1+\epsilon\,ip_2)\,
 e^{-a (p_\perp+v_\perp/2)^2-b (p_\perp-v_\perp/2)^2} \ = \
-\,\frac{1}{8\pi}\,(v_1+\epsilon\,iv_2)\, \frac{(a - b)\;}{(a+b)^2}\;
e^{-\frac{a b}{a+b}\;v_\perp^2}\ .
\end{equation}

On the other hand, the integrals over $v_\perp$ can be obtained taking into
account the following useful relations. Defining
\begin{equation}
{\cal D}_k(x_1,q_2,v_\perp) \ = \ D_k(\sqrt{2B_e}\,x'_1+\sqrt{2/B_e}\,q_2)\;
e^{iv_1(x_1-x'_1)}\,\Big|_{x_1'=-x_1+2(v_2-q_2)/B_e}\ \ ,
\end{equation}
one has
\begin{eqnarray}
\int_{v_\perp}{\cal D}_k(x_1,q_2,v_\perp)\;
e^{-\gamma v_\perp^2} & = & \frac{B_e}{4\pi}\,\frac{(1-\gamma B_e)^k}{(1+\gamma B_e)^{k+1}}
\,D_k(\sqrt{2B_e}\,x_1+\sqrt{2/B_e}\,q_2)\ ,
\nonumber \\
\int_{v_\perp}(v_1+\epsilon\,iv_2)\,{\cal D}_k(x_1,q_2,v_\perp)\;
e^{-\gamma v_\perp^2} & = & \frac{i\,\epsilon\sqrt{2}\,B_e^{3/2}}{4\pi}
\,\Big(\frac{1-\gamma B_e}{1+\gamma B_e}\Big)^{k+\epsilon}
\frac{k^{(1-\epsilon)/2}}{(1+\gamma B_e)\,(1-\epsilon\gamma B_e)}\, \times
\nonumber \\
& &
D_{k+\epsilon}(\sqrt{2B_e}\,x_1+\sqrt{2/B_e}\,q_2)\ ,
\nonumber \\
\int_{v_\perp}\, v_\perp^2\,{\cal D}_k(x_1,q_2,v_\perp)\;
e^{-\gamma v_\perp^2}
& = & \frac{B_e^2}{4\pi}\,\frac{(1-\gamma B_e)^{k-1}}{(1+\gamma B_e)^{k+2}}\,(1-\gamma B_e+2k)
\,\times \nonumber \\
& & D_k(\sqrt{2B_e}\,x_1+\sqrt{2/B_e}\,q_2)\ .
\end{eqnarray}


\begin{thebibliography}{199}

\bibitem{Kharzeev:2012ph}
D.~E.~Kharzeev, K.~Landsteiner, A.~Schmitt and H.~U.~Yee,
Lect.\ Notes Phys.\  {\bf 871}, 1 (2013).

\bibitem{Andersen:2014xxa}
  J.~O.~Andersen, W.~R.~Naylor and A.~Tranberg,
  Rev.\ Mod.\ Phys.\  {\bf 88}, 025001 (2016).

\bibitem{Miransky:2015ava}
  V.~A.~Miransky and I.~A.~Shovkovy,
  Phys.\ Rep.\  {\bf 576}, 1 (2015).

\bibitem{Grasso:2000wj}
  D.~Grasso and H.~R.~Rubinstein,
  Phys.\ Rep.\  {\bf 348}, 163 (2001).

\bibitem{HIC}
D.~E.~Kharzeev, L.~D.~McLerran and H.~J.~Warringa, Nucl.\ Phys.\ A {\bf 803}, 227 (2008);
V.~Skokov, A. Y.~Illarionov, and V.~Toneev, Int. J. Mod. Phys. A {\bf 24}, 5925 (2009);
V.~Voronyuk, V.~Toneev, W.~Cassing, E.~Bratkovskaya, V.~Konchakovski, and S.~Voloshin,
Phys. Rev. C {\bf 83}, 054911 (2011).

\bibitem{duncan}
R.~C.~Duncan and C.~Thompson, Astrophys. J. {\bf 392}, L9 (1992); C.~Kouveliotou {\it et al}.,
Nature (London) {\bf 393}, 235 (1998).

\bibitem{Fayazbakhsh:2012vr}
  S.~Fayazbakhsh, S.~Sadeghian and N.~Sadooghi,
  Phys.\ Rev.\ D {\bf 86}, 085042 (2012).

\bibitem{Fayazbakhsh:2013cha}
  S.~Fayazbakhsh and N.~Sadooghi,
  Phys.\ Rev.\ D {\bf 88}, 065030 (2013).

\bibitem{Avancini:2015ady}
  S.~S.~Avancini, W.~R.~Tavares and M.~B.~Pinto,
  Phys.\ Rev.\ D {\bf 93}, 014010 (2016).

\bibitem{Avancini:2016fgq}
  S.~S.~Avancini, R.~L.~S.~Farias, M.~Benghi Pinto, W.~R.~Tavares and V.~S.~Timoteo,
  Phys.\ Lett.\ B {\bf 767}, 247 (2017).

\bibitem{Mao:2017wmq}
  S.~Mao and Y.~Wang,
  Phys.\ Rev.\ D {\bf 96}, 034004 (2017).

\bibitem{Zhang:2016qrl}
  R.~Zhang, W.~j.~Fu and Y.~x.~Liu,
  Eur.\ Phys.\ J.\ C {\bf 76}, 307 (2016).

\bibitem{GomezDumm:2017jij}
  D.~Gomez Dumm, M.~F.~I.~Villafa\~ne and N.~N.~Scoccola,
  Phys.\ Rev.\ D {\bf 97}, 034025 (2018).

\bibitem{Wang:2017vtn}
  Z.~Wang and P.~Zhuang,
  Phys.\ Rev.\ D {\bf 97}, 034026 (2018).

\bibitem{Liu:2018zag}
  H.~Liu, X.~Wang, L.~Yu and M.~Huang,
  Phys.\ Rev.\ D {\bf 97}, 076008 (2018).

\bibitem{Coppola:2018vkw}
  M.~Coppola, D.~Gomez Dumm and N.~N.~Scoccola,
  Phys.\ Lett.\ B {\bf 782}, 155 (2018).

\bibitem{Mao:2018dqe}
  S.~Mao,
  Phys.\ Rev.\ D {\bf 99}, 056005 (2019).

\bibitem{Avancini:2018svs}
  S.~S.~Avancini, R.~L.~S.~Farias and W.~R.~Tavares,
  Phys.\ Rev.\ D {\bf 99}, 056009 (2019).

\bibitem{Ayala:2018zat}
  A.~Ayala, R.~L.~S.~Farias, S.~Hern\'andez-Ortiz, L.~A.~Hern\'andez, D.~M.~Paret and R.~Zamora,
  Phys.\ Rev.\ D {\bf 98}, 114008 (2018).

\bibitem{Kamikado:2013pya}
  K.~Kamikado and T.~Kanazawa,
  JHEP {\bf 1403}, 009 (2014).

\bibitem{Agasian:2001ym}
  N.~O.~Agasian and I.~A.~Shushpanov,
  JHEP {\bf 0110}, 006 (2001).

\bibitem{Andersen:2012zc}
  J.~O.~Andersen,
  JHEP {\bf 1210}, 005 (2012).

\bibitem{Colucci:2013zoa}
  G.~Colucci, E.~S.~Fraga and A.~Sedrakian,
  Phys.\ Lett.\ B {\bf 728}, 19 (2014).

\bibitem{Orlovsky:2013wjd}
  V.~D.~Orlovsky and Y.~A.~Simonov,
  JHEP {\bf 1309}, 136 (2013).

\bibitem{Andreichikov:2016ayj}
  M.~A.~Andreichikov, B.~O.~Kerbikov, E.~V.~Luschevskaya, Y.~A.~Simonov and O.~E.~Solovjeva,
  JHEP {\bf 1705}, 007 (2017).

\bibitem{Simonov:2015xta}
  Y.~A.~Simonov,
  Phys.\ Atom.\ Nucl.\  {\bf 79}, 455 (2016)
  [Yad.\ Fiz.\  {\bf 79}, 277 (2016)].

\bibitem{Andreichikov:2018wrc}
M.~A.~Andreichikov and Y.~A.~Simonov,
  Eur.\ Phys.\ J.\ C {\bf 78}, 902 (2018).

\bibitem{Dominguez:2018njv}
  C.~A.~Dominguez, M.~Loewe and C.~Villavicencio,
  Phys.\ Rev.\ D {\bf 98}, 034015 (2018).

\bibitem{Bali:2011qj}
  G.~S.~Bali, F.~Bruckmann, G.~Endrodi, Z.~Fodor, S.~D.~Katz, S.~Krieg, A.~Schafer and K.~K.~Szabo,
  JHEP {\bf 1202}, 044 (2012).

\bibitem{Hidaka:2012mz}
  Y.~Hidaka and A.~Yamamoto,
  Phys.\ Rev.\ D {\bf 87}, 094502 (2013).

\bibitem{Luschevskaya:2014lga}
  E.~V.~Luschevskaya, O.~E.~Solovjeva, O.~A.~Kochetkov and O.~V.~Teryaev,
  Nucl.\ Phys.\ B {\bf 898}, 627 (2015).

\bibitem{Bali:2015vua}
  B.~B.~Brandt, G.~Bali, G.~Endrodi and B.~Glaessle,
  PoS LATTICE {\bf 2015}, 265 (2016).

\bibitem{Bali:2017ian}
  G.~S.~Bali, B.~B.~Brandt, G.~Endrodi and B.~Glaessle,
  Phys.\ Rev.\ D {\bf 97}, 034505 (2018).

\bibitem{Vogl:1991qt}
  U.~Vogl and W.~Weise,
  Prog.\ Part.\ Nucl.\ Phys.\ {\bf 27}, 195 (1991).

\bibitem{Klevansky:1992qe}
  S.~P.~Klevansky,
  Rev.\ Mod.\ Phys.\  {\bf 64}, 649 (1992).

\bibitem{Hatsuda:1994pi}
  T.~Hatsuda and T.~Kunihiro,
  Phys.\ Rep.\  {\bf 247}, 221 (1994).

\bibitem{Schwinger:1951nm}
  J.~S.~Schwinger,
  Phys.\ Rev.\  {\bf 82}, 664 (1951).

\bibitem{Ritus:1978cj}
  V.~I.~Ritus,
  Sov.\ Phys.\ JETP {\bf 48}, 788 (1978).

\bibitem{Avancini:2019wed}
  S.~S.~Avancini, R.~L.~S.~Farias, N.~N.~Scoccola and W.~R.~Tavares,
  Phys.\ Rev.\ D {\bf 99}, 116002 (2019).

\bibitem{Bali:2018sey}
  G.~S.~Bali, B.~B.~Brandt, G.~Endr{\H o}di and B.~Gl{\"a}{\ss}le,
  Phys.\ Rev.\ Lett.\  {\bf 121}, 072001 (2018).

\bibitem{Coppola:2018ygv}
  M.~Coppola, D.~Gomez Dumm, S.~Noguera and N.~N.~Scoccola,
  Phys.\ Rev.\ D {\bf 99}, 054031 (2019).

\bibitem{Menezes:2008qt}
  D.~P.~Menezes, M.~Benghi Pinto, S.~S.~Avancini, A.~Perez Martinez and C.~Providencia,
  Phys.\ Rev.\ C {\bf 79}, 035807 (2009).

\bibitem{Allen:2015paa}
  P.~G.~Allen, A.~G.~Grunfeld and N.~N.~Scoccola,
  Phys.\ Rev.\ D {\bf 92}, 074041 (2015).


\bibitem{Miransky:2002rp}
  V.~A.~Miransky and I.~A.~Shovkovy,
  Phys.\ Rev.\ D {\bf 66}, 045006 (2002).


\bibitem{WZW}
  J.~Wess and B.~Zumino,
  Phys.\ Lett.\  {\bf 37B}, 95 (1971);
  E.~Witten,
  Nucl.\ Phys.\ B {\bf 223}, 422 (1983).


\bibitem{Gusynin:1995nb}
  V.~P.~Gusynin, V.~A.~Miransky and I.~A.~Shovkovy,
  Nucl.\ Phys.\ B {\bf 462}, 249 (1996).

\bibitem{Fukushima:2012kc}
  K.~Fukushima and Y.~Hidaka,
  Phys.\ Rev.\ Lett.\  {\bf 110}, 031601 (2013).

\end{thebibliography}
\end{document}